\documentclass[12pt, draftclsnofoot, onecolumn]{IEEEtran}
\usepackage{graphicx}
\usepackage{subfigure}
\usepackage{float}
\usepackage{subfloat}
\usepackage{amsmath}
\usepackage{amssymb}
\usepackage{amsthm}
\usepackage{epstopdf}
\usepackage{cases}
\usepackage{color}
\usepackage{makecell}
\usepackage{cite}
\usepackage{algorithmic}
\usepackage[ruled]{algorithm2e}
\usepackage{caption}
\usepackage{mathtools}

\newtheorem{Theo}{Theorem}

\usepackage[normalem]{ulem}

\ifCLASSINFOpdf
\else
\fi
\begin{document}

%
\title{Multiple Signal Classification Based Joint Communication and Sensing System}
%
%
%

\author{Xu Chen,~\IEEEmembership{Member,~IEEE,}
	Zhiyong Feng,~\IEEEmembership{Senior Member,~IEEE,}\\
	Zhiqing Wei,~\IEEEmembership{Member,~IEEE,}
	Xin Yuan,~\IEEEmembership{Member,~IEEE,}\\
	Ping Zhang,~\IEEEmembership{Fellow,~IEEE,}
	J. Andrew Zhang,~\IEEEmembership{Senior Member,~IEEE,}\\
	and Heng Yang,~\IEEEmembership{Member,~IEEE}

	\thanks{This work is supported by the National Key Research and Development Program of	China under Grants \{2020YFA0711300, 2020YFA0711302, and 2020YFA0711303\}, and the National Natural Science Foundation of China under Grants \{61941102, 61790553\}, and BUPT Excellent Ph.D. Students Foundation under grant CX2021110.}
	\thanks{X. Chen, Z. Feng, Z. Wei, and H. Yang are with Beijing University of Posts and Telecommunications, Key Laboratory of Universal Wireless Communications, Ministry of Education, Beijing 100876, P. R. China (Email:\{chenxu96330, fengzy, weizhiqing, yangheng\}@bupt.edu.cn).}
	\thanks{X. Yuan is with Commonwealth Scientific and Industrial Research Organization (CSIRO), Australia (email: Xin.Yuan@data61.csiro.au).}
	\thanks{P. Zhang is with Beijing University of Posts and Telecommunications, State Key Laboratory of Networking and Switching Technology, Beijing 100876, P. R. China (Email: pzhang@bupt.edu.cn).}
	\thanks{J. A. Zhang is with the Global Big Data Technologies Centre, University of Technology Sydney, Sydney, NSW, Australia (Email: Andrew.Zhang@uts.edu.au).}
	
	\thanks{Corresponding author: Zhiyong Feng, Zhiqing Wei.}
}

%
%

\markboth{}%
{Shell \MakeLowercase{\textit{et al.}}: Bare Demo of IEEEtran.cls for IEEE Journals}
%



\maketitle

\pagestyle{empty}  
\thispagestyle{empty} 

\newcounter{mytempeqncnt}
\setcounter{mytempeqncnt}{\value{equation}}
\begin{abstract}


\color{blue} Joint communication and sensing (JCS) has become a promising technology for mobile networks because of its higher spectrum and energy efficiency. Up to now, the prevalent fast Fourier transform (FFT)-based sensing method for mobile JCS networks is on-grid based, and the grid interval determines the resolution. Because the mobile network usually has limited consecutive OFDM symbols in a downlink (DL) time slot, the sensing accuracy is restricted by the limited resolution, especially for velocity estimation. In this paper, we propose a multiple signal classification (MUSIC)-based JCS system that can achieve higher sensing accuracy for the angle of arrival, range, and velocity estimation, compared with the traditional FFT-based JCS method. We further propose a JCS channel state information (CSI) enhancement method by leveraging the JCS sensing results. Finally, we derive a theoretical lower bound for sensing mean square error (MSE) by using perturbation analysis. Simulation results show that in terms of the sensing MSE performance, the proposed MUSIC-based JCS outperforms the FFT-based one by more than 20 dB. Moreover, the bit error rate (BER) of communication demodulation using the proposed JCS CSI enhancement method is significantly reduced compared with communication using the originally estimated CSI.

\end{abstract}


%
\IEEEpeerreviewmaketitle

\section{Introduction}
%
%
%
%
\subsection{Background and Motivations}
{\color{blue} Wireless communication and sensing are both indispensable for critical machine-type applications, e.g., the 5th generation (5G) and the future 6th generation (6G) networks~\cite{Saad2020,You2020,ITUITShandbook}. Nevertheless, the proliferation of wireless sensing and communication infrastructures and devices will result in severe spectrum congestion problems~\cite{liu2020joint}. Joint communication and sensing (JCS) has emerged as one of the most promising 6G key techniques due to its potential in improving spectrum and energy efficiency. It aims to achieve wireless sensing and communication simultaneously using unified spectrum and transceivers,  sharing the same transmitted signals~\cite{Zhang2019JCRS}.}

\subsection{Related Works}
{\color{blue} Since orthogonal frequency-division multiplexing (OFDM) is the most popular physical-layer signal solution for broadband wireless networks, the JCS techniques based on OFDM signals have been widely researched. Sturm \textit{et al}.~\cite{Sturm2011Waveform} proposed a fast Fourier transform (FFT)-based frequency-domain OFDM JCAS signal processing method, realizing both active range estimation and communication. By utilizing the FFT-based JCS signal processing method, Zhang \textit{et al.}~\cite{Zhang2019JCRS} proposed a practical OFDM JCS system based on the time-division-duplex (TDD) mobile network, which is suitable for downlink (DL) echo sensing. In~\cite{Kumari2018WIFI}, the authors proposed an IEEE 802.11ad-based OFDM JCS vehicle-to-vehicle (V2V) system exploiting the preamble of a single-carrier physical layer frame to achieve V2V communication and full-duplex radar in the 60 GHz band. In~\cite{Chen2021CDOFDM}, the authors proposed a code-division OFDM JCS system by introducing code-division multiplex into FFT-based OFDM JCS processing to improve the JCS sensing performance. As pointed out in~\cite{Andrew2021PMN}, the full-duplex (FD) is the critical enabler for implementing DL JCS, which can simultaneously transmit JCAS signals and receive reflections. 
Seyed Ali~\textit{et al.}~\cite{IBFDJCR} realized an FD JCS platform that detects targets while communicating with another node by canceling the self-leakage interference with analog and digital self-leakage canceler.
	
Despite the above studies, there is a huge obstacle to utilizing the FFT-based OFDM JCS method in real applications. This method has to use consecutive OFDM subcarriers and symbols to estimate the range and velocity on the fixed grid, while the grid interval determines the resolution. Therefore, the range and velocity resolutions are determined by the number of used subcarriers and OFDM symbols, respectively. Thus, for mobile networks that typically have limited subcarriers and OFDM symbols, e.g., 14 OFDM symbols in each DL time slot, the sensing accuracy, especially the velocity accuracy, is largely restricted.}  Besides, in~\cite{Fang2020}, the author showed that overlapped interference deteriorates sensing performance in a networking situation. Therefore, it is also important for a sensing method that can still work effectively under a low signal to interference plus noise ratio (SINR).

\subsection{Our Contributions}
{\color{blue} To resolve the aforementioned problems, we propose a multiple signal classification (MUSIC)-based sensing scheme for OFDM JCS systems that can achieve accurate estimation of the angle of arrival (AoA), range, and velocity, adapting to various OFDM communication signals with limited OFDM subcarriers and symbols. We also propose a JCS channel state information (CSI) enhancement method that exploits the JCS sensing results for refining CSI estimation with a Kalman filter. Furthermore, we provide some theoretical lower bound of mean square error (MSE) for the proposed MUSIC-based JCS sensing algorithms.}

The main contributions of this paper are summarized as follows. 
\begin{itemize}
	\item[1.] {\color{blue} We propose a novel MUSIC-based JCS range and velocity estimation scheme, which consists of expanded two-dimensional (2D) MUSIC algorithms and two-step descent searching algorithms. The proposed scheme can use communication signals to achieve accurate range and velocity estimation.}
	
	\item[2.] {\color{blue} We propose a JCS CSI enhancement method based on the Kalman filter, which exploits the JCS sensing parameters to construct the state transfer model and refines the CSI estimation using the JCS sensing results. This method can improve the bit error rate (BER) in the case of imperfect CSI.}
	
	\item[3.] {\color{blue} We derive the theoretical MSEs for the proposed MUSIC-based JCS range and velocity estimation scheme using perturbation analysis. The theoretical MSEs of range and velocity estimation match the simulation MSEs well in the high SINR regime.}
	
	\item[4.] {\color{blue} Extensive simulations are conducted to validate the proposed JCS sensing and CSI enhancement schemes and the theoretical MSEs. The results show that the proposed sensing scheme outperforms the conventional 2D-FFT method in terms of range and Doppler estimation MSEs by more than 20 dB, and the JCS CSI enhancement method can significantly improve communication performance.}
\end{itemize}

\subsection{Organization and Notations}
The remaining parts of this paper are organized as follows. 
In Section \ref{sec:system-model}, we describe the DL JCS model and transmitting signal model, and propose the JCS channel model. 
Section \ref{sec:JCS Downlink signal processing} proposes the MUSIC-based JCS AoA, range and velocity estimation method.
Section \ref{sec:JCS performance} provides a theoretical analysis of the proposed estimation method.
In Section \ref{sec:JCS result}, the simulation results are presented. 
Section \ref{sec:conclusion} concludes this paper.

\textbf{Notations:} Bold uppercase letters denote matrices (e.g., $\textbf{M}$); bold lowercase letters denote column vectors (e.g., $\textbf{v}$); scalars are denoted by normal font (e.g., $\gamma$); the entries of vectors or matrices are referred to with brackets, for instance, the $q$th entry of vector $\textbf{v}$ is $[\textbf{v}]_{q}$, and the entry of the matrix $\textbf{M}$ at the $m$th row and $q$th column is ${[\textbf{M}]_{n,m}}$; $\left(\cdot\right)^H$, $\left(\cdot\right)^{*}$ and $\left(\cdot\right)^T$ denote Hermitian transpose, complex conjugate and transpose, respectively; ${\left\| {\mathbf{v}}_{k}  \right\|_l}$ represents the $l$-norm of ${\mathbf{v}}_{k}$; $E\left( \cdot \right)$ represents the expectation of random variables; {\color{blue} ${\bf M}_1 \in \mathbb{C}^{M\times N}$ and ${\bf M}_2 \in \mathbb{R}^{M\times N}$ represent that ${\bf M}_1$ and ${\bf M}_2$ are ${M\times N}$ complex-value and real-value matrices, respectively, and $v \sim \mathcal{CN}(m,\sigma^2)$ means $v$ follows a complex Gaussian distribution with mean $m$ and variance $\sigma^2$.
}

\section{System Model}\label{sec:system-model}
\subsection{DL JCS Model} \label{subsec:The Downlink JCS Model}
\begin{figure}[!t]
	\captionsetup{width=0.47\textwidth}
	\centering
	\begin{minipage}[t]{0.52\linewidth}
		\centering
		\includegraphics[width=\columnwidth]{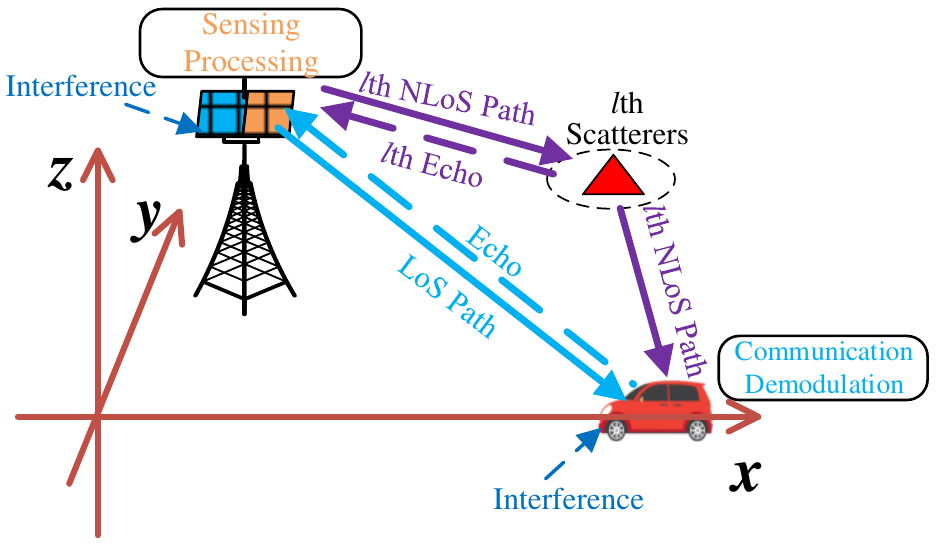}
		\caption{DL JCS Scenario.}
		\label{fig: Downlink JCS Model}
	\end{minipage}
	\begin{minipage}[t]{0.47\linewidth}
		\centering
		\includegraphics[width=\columnwidth]{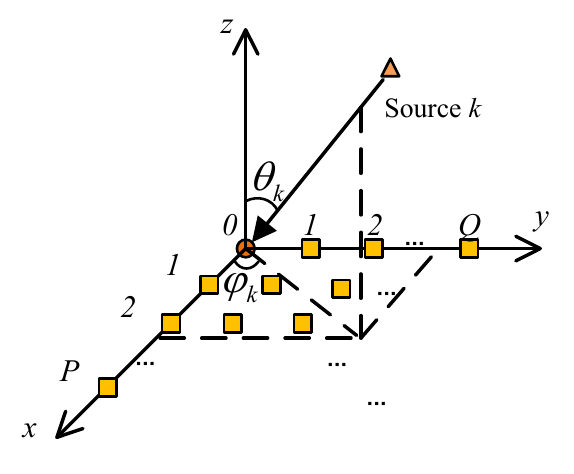}
		\caption{UPA model.}
		\label{fig: UPA model}
	\end{minipage}%
\end{figure}
As shown in Fig.~\ref{fig: Downlink JCS Model}, we consider the DL JCS process between the BS and the machine-type user equipment (MUE), such as a road-side infrastructure and a vehicle. Millimeter-wave (mmWave) signal is considered for DL JCS. It is particularly suitable for JCS given its potential high resolution. {\color{blue} The BS and MUEs are equipped with uniform plane arrays (UPAs). The BS is equipped with two spatially well-separated UPAs and a self-leakage canceler to realize the FD capability, as detailed in~\cite{IBFDJCR}. Therefore, the self-leakage between arrays is ignored and not considered in the signal model in this paper. One BS array is used for transmitting the DL JCS signal, and the other is used for consistently receiving echoes of the JCS signal. MUE receives the JCS signal to demodulate the communication data, while BS receives the echoes to estimate the AoAs, ranges, and velocities. Moreover, we consider that both BS and MUE  receive the superimposed co-channel interference from multiple reflected interference sources (ISs). The MUE is equipped with one UPA for receiving the communication signal. The array sizes of the BS and MUEs are ${P_t} \times {Q_t}$ and ${P_r} \times {Q_r}$, respectively.
}

\subsection{UPA Model} \label{subsec:UPA model}
Fig.~\ref{fig: UPA model} demonstrates the model of UPAs. The uniform interval between the neighboring antenna elements is denoted by $d_a$. The size of the UPA is denoted by ${P} \times {Q}$. The two-dimensional (2D) AoA for receiving or the AoD for transmitting the $k$th far-field signal is ${{\bf{p}}_k} = {\left( {{\varphi _k},{\theta _k}} \right)^T}$, where ${\varphi _k}$ is the azimuth angle, and ${\theta _k}$ is the elevation angle. We use ${A_{p,q}}$ to denote the
($p$,$q$)th antenna element, and ${A_{0,0}}$ to represent the reference antenna element. Then, the phase difference between ${A_{p,q}}$ and ${A_{0,0}}$ is expressed as
\begin{equation}\label{equ:phase_difference}
	{a_{p,q}} \!\left( {{{\bf{p}}}} \right) \! =\! \exp \!\left[  \!{ - j \!\frac{{2\pi }}{\lambda }{d_a}  \!\!\left( {p\cos {\varphi} \sin {\theta}  \!+ \! q\sin {\varphi} \sin {\theta}} \right)} \!\right],
\end{equation}
where $\lambda = c/f_c$ is the wavelength of the carrier, $c$ is the speed of light in vacuum, and $f_c$ is the carrier frequency. 

The steering vector for the array is
\begin{equation}\label{equ:steeringVec}
	{\bf{a}}\left( {{{\bf{p}}_k}} \right) = \left[ {{a_{p,q}}\left( {{{\bf{p}}_k}} \right)} \right]\left| {_{p = 0,1,...,P - 1;q = 0,1,...,Q - 1}}\right.,
\end{equation}
where ${\bf{a}}\left( {{{\bf{p}}_k}} \right)$ is a ${P}{Q} \times 1$ vector, and ${\left. {\left[ {{v_{p,q}}} \right]} \right|_{(p,q) \in {\bf{S}}1 \times {\bf{S}}2}}$ denotes the vector stacked by ${v_{p,q}}$ satisfying $p\in{\bf{S}}1$ and $q\in{\bf{S}}2$. 

The steering matrix for $K$ far-field signals is then represented as 
\begin{equation}\label{equ:steering Matrix}
	{\bf{A}} = \left[ {{\bf{a}}\left( {{{\bf{p}}_1}} \right),{\bf{a}}\left( {{{\bf{p}}_2}} \right),...,{\bf{a}}\left( {{{\bf{p}}_K}} \right)} \right],
\end{equation}
which is a matrix of dimension ${P}{Q} \times K$.

\subsection{DL JCS Signal and Channel Model} \label{subsec:uplink_signal model}

In this paper, we consider the JCS system using OFDM-based signals. The transmitting signal is 
\begin{equation}\label{equ:OFDM transmit signal}
	{s_D}\left( t \right)\! =\!\! \sum\limits_{m = 0}^{M_s \!-\! 1}\!{\sum\limits_{n = 0}^{N_c \!-\! 1} \!\!\!{\sqrt {P_t} d_{n,m}{e^{j2\pi \left( {{f_c} + n\Delta {f}} \right)t}}} } {\mathop{\rm Rect}\nolimits} (\frac{{t - mT}}{{T}}),
\end{equation}
where $P_t$ is the DL transmit power, $d_{n,m}$ is the $m$th baseband OFDM symbol of the $n$th subcarrier, $\Delta {f}$ is the subcarrier interval, $T = T_s + {T_g}$, $T_s = \frac{1}{{\Delta {f}}}$ is the duration of OFDM symbol, ${T_g}$ is the guard interval, $M_s$ and $N_c$ are the number of OFDM symbols and subcarriers, respectively, and ${\rm Rect}\left( t/T\right)$ is the rectangular window function of duration $T$. {\color{blue} When the DL preamble signal for beam alignment and CSI estimation is transmitted, $d_{n,m}$ is replaced by the preamble symbols, denoted by $\bar d_{n,m}$, which is known and deterministic to both BS and MUE. When the DL data signal is transmitted, $d_{n,m} \in {\Theta _{QAM}}$ is a random symbol, where ${\Theta _{QAM}}$ is the constellation of quadrature amplitude modulation (QAM). Note that $d_{n,m}$ is known to BS but unknown to MUE.}

{\color{blue}
Next, we present the JCS channel model. 
As illustrated in Fig.~\ref{fig: Downlink JCS Model}, the DL JCS channel comprises a communication channel and an echo sensing channel. 
\begin{itemize}
	\item The JCS communication channel consists of a line-of-sight (LoS) path and several non-line-of-sight (NLoS) scattering paths.
	\item The JCS echo sensing channel consists of the echo path from MUE as a scatterer, and the echo paths from other scatterers which may or may not contribute to the communication channel. Since the signals after multiple reflections are much smaller than those with only one reflection, we only consider echoes directly reflected from scatterers.
\end{itemize}
}

{\color{blue}
Then, the JCS sensing echo and communication channels at the $n$th subcarrier of the $m$th OFDM symbol are defined as~\cite{Chen2021CDOFDM, Zhang2019JCRS}
\begin{equation}\label{equ:general_JCS_channel}
	{{\bf{H}}_{i,n,m}}{\rm{ = }}\sum\limits_{l = 0}^{L - 1} {\left[ {{\alpha _{i,n,m,l}}{\bf{a}}({\bf{p}}_{RX,l}^i){\bf{a}}({{\bf{p}}_{TX,l}})} \right]},
\end{equation}
where ${\bf{p}}_{TX,l}$ is the AoD of BS's JCS transceivers, ${{\bf{a}}( {{\bf{p}}_{TX,l}} )} \in \mathbb{C}^{{P_t}{Q_t} \times 1}$ is the corresponding transmit steering vectors as given in \eqref{equ:steeringVec}, $L$ is the number of scatterers, $l = 0$ is for the direct path between BS and MUE, $l = 1, \cdots, L-1$ is for the reflected paths involved the $l$th scatterer. Moreover, $i = S$ and $i = C$ represent the echo sensing and communication channels, respectively; ${\bf{p}}_{RX,l}^S$ and ${\bf{p}}_{RX,l}^C$ are the AoAs of BS's echo receiver and the MUE's communication receiver, respectively; and ${\alpha _{S,n,m,l}}$ and ${\alpha _{C,n,m,l}}$ are the channel fading for the $l$th sensing echo path and communication path, respectively.
}

\subsubsection{JCS Echo Sensing Channel}
{\color{blue} When $i = S$, ${{\bf{a}}( {\bf{p}}_{RX,l}^{S} )}\in \mathbb{C}^{{P_t}{Q_t} \times 1}$ is the receive steering vector for the $l$th echo sensing path, as given in \eqref{equ:steeringVec}. Since the mmWave array is typically small, ${\bf{p}}_{RX,l}^{S} = {{\bf{p}}_{TX,l}}$. Moreover, ${\alpha _{S,n,m,l}}$ is the fading factor for the $l$th echo (when $l = 0$, MUE acts as a scatterer), which is given by
	\begin{equation}\label{equ:alpha_S}
		{\alpha _{S,n,m,l}} = {b_{S,l}}{e^{j2\pi m{T}{f_{s,l,1}}}}{e^{ - j2\pi n\Delta {f} {{\tau _{s,l}}} }},
	\end{equation}
	where ${f_{s,0,1}} = \frac{{2{v_{r,0,1}}}}{\lambda }$ and ${\tau _{s,0}} = \frac{{2{d_{0,1}}}}{c}$ are the echo Doppler frequency shifts and time delay between BS and MUE, with $v_{r,0,1}$ and $d_{0,1}$ being the corresponding radial relative velocity and the distance, respectively; ${f_{s,l,1}} = \frac{{2{v_{r,l,1}}}}{\lambda }$ and ${\tau _{s,l}} = \frac{{2{d_{l,1}}}}{c}$ are the echo Doppler frequency shifts and time delay between BS and the $l$th scatterer, with $v_{r,l,1}$ and ${d_{l,1}}$ being the corresponding radial relative velocity and distance, respectively. Moreover, ${b_{S,l}} = \sqrt {\frac{{{\lambda ^2}}}{{{{\left( {4\pi } \right)}^3}{d_{l,1}}^4}}} {\beta _{S,l}}$, and ${\beta _{S,l}}$ is the random reflection fading factor of the $l$th scatterer, following the complex Gaussian distribution with zero mean and variance $\sigma _{S\beta ,l}^2$. 
}

\subsubsection{JCS Communication Channel}
When $i = C$, ${\bf{a}}( {{\bf{p}}^{C}_{RX,l}} ) \in \mathbb{C}^{{P_r}{Q_r} \times 1}$ is the receive steering vector for the $l$th communication path, as given in \eqref{equ:steeringVec}. Moreover, ${\alpha _{C,n,m,l}}$ is the fading factor for the $l$th path, and is expressed as
\begin{equation}\label{equ:alpha_C}
	{\alpha _{C,n,m,l}} \!= \! \left\{ \!\!\! \begin{array}{l}
		{b_{C,0}}{e^{j2\pi  mT{{f_{c,d,0}}} }}{e^{ - j2\pi n\Delta {f} {{\tau _{c,0}}} }},l = 0\\
		\!\!\!\begin{array}{l}
			{b_{C,l}}{e^{j2\pi ( {{f_{d,l,1}} + {f_{d,l,2}}} )mT}}
			{e^{ - j2\pi n\Delta {f}( {{\tau _{c,l,1}} + {\tau _{c,l,2}}} )}}
		\end{array},l > 0
	\end{array} \right.,
\end{equation}
where ${f_{c,d,0}} = \frac{{{v_{r,0,1}}}}{\lambda }$ and ${\tau _{c,0}} = \frac{{{d_{0,1}}}}{c}$ are the Doppler frequency shift and time delay of the LoS path; ${f_{d,l,1}} = \frac{{{v_{r,l,1}}}}{\lambda }$, ${f_{d,l,2}} = \frac{{{v_{r,l,2}}}}{\lambda }$, ${\tau _{c,l,1}} = \frac{{{d_{l,1}}}}{c}$ and ${\tau _{c,l,2}} = \frac{{{d_{l,2}}}}{c}$ are the Doppler frequency shifts and time delay between BS and scatterer, and between the scatterer and MUE of the $l$th NLoS path, respectively, with $v_{r,l,2}$ and ${d_{l,2}}$ being the radial relative velocity and distance between the $l$th scatterer and MUE, respectively; ${b_{C,0}} = \sqrt {\frac{{{\lambda ^2}}}{{{{(4\pi {d_0})}^2}}}}$ is the propagation loss of the LoS path, and ${b_{C,l}} = \sqrt {\frac{{{\lambda ^2}}}{{{{\left( {4\pi } \right)}^3}{d_{l,1}}^2{d_{l,2}}^2}}}   {\beta _{C,l}}$ is the path fading factor of the $l$th NLoS path with ${\beta _{C,l}}$ being the scattering factor of the $l$th scatterer. 
{\color{blue} Here, ${\beta _{C,l}}$ is the random reflecting factor of the scatterer in the $l$th path, which is assumed to follow the complex Gaussian distribution with zero mean and variance $\sigma _{C\beta ,l}^2$.}
Due to the existence of ${b_{C,l}}$, the LoS path is much stronger than the NLoS path for mmWave.

{\color{blue}
Note that ${\bf{H}}_{C,n,m}$ is unknown and needs to be estimated by utilizing the DL preambles, $\bar d_{n,m}$. The parameters of ${\bf{H}}_{S,n,m}$ are unknown, and BS has to estimate the AoA, range and Doppler in ${\bf{H}}_{S,n,m}$. Since BS acts as both the sensing transmitter and receiver, both $d_{n,m}$ and $\bar d_{n,m}$ can be used for DL sensing. Moreover, $l = 0$ represents a special path, for which the echo time delay and Doppler are twice of those in the communication channel, which is the theoretical basis for the JCS CSI enhancement method to be introduced in Section \ref{sec:JCS_comm}. 
}

\begin{figure}[!t]
	\centering
	\includegraphics[width=0.40\textheight]{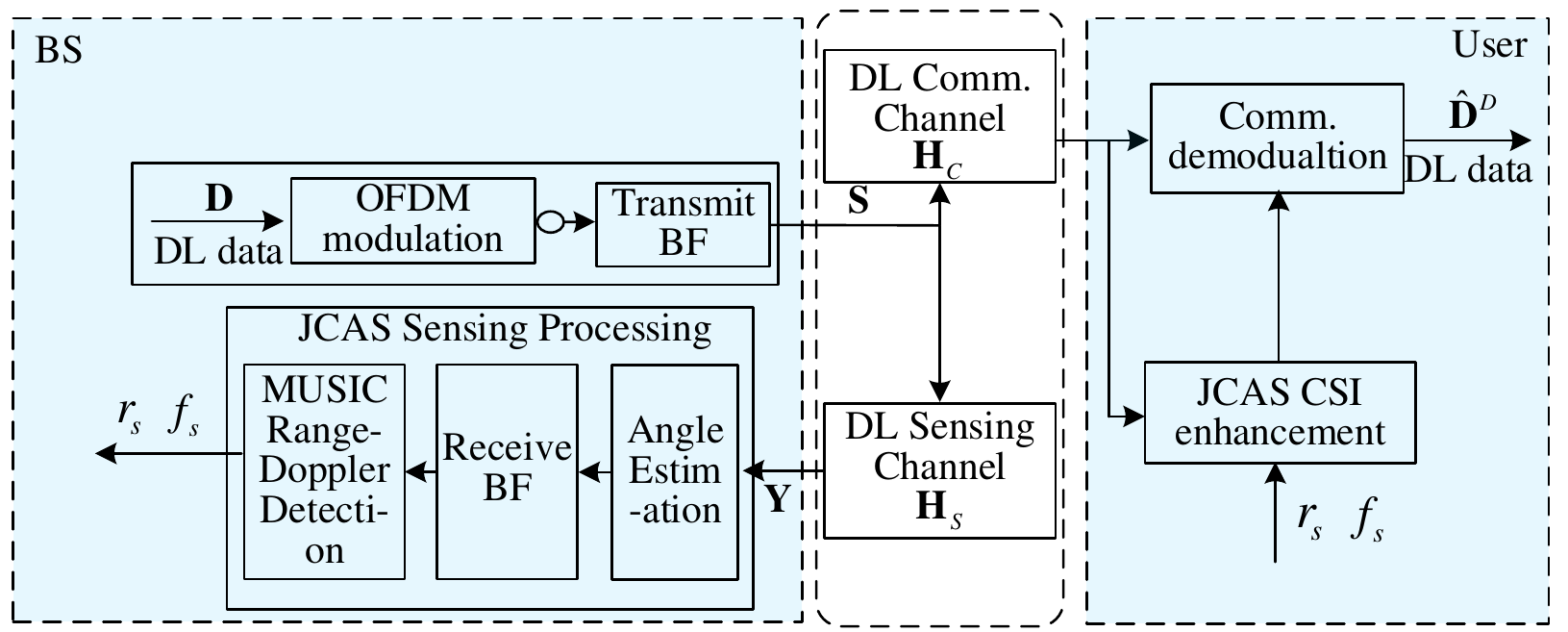}%
	\DeclareGraphicsExtensions.
	\caption{DL JCS signal processing diagram.}
	\label{fig: DL_JCS_signal_processing}
\end{figure}

\subsection{JCS Received Signal Model} \label{sec:JCS Signal_reception}
In this subsection, we present the expressions for DL JCS received signals.

{\color{blue} 
\subsubsection{DL Communication Received Signal} 
The frequency-domain DL communication signal received by MUE at the $m$th OFDM symbol of the $n$th subcarrier is expressed as 
\begin{equation}\label{equ:y_DC}
	y_{C,n,m} = \sqrt {P_t} d_{n,m}{\left( {{\bf{w}}_{RX}} \right)^H}{\bf{H}}_{C,n,m}{\bf{w}}_{TX} + n^{X}_{C,n,m},
\end{equation}
where ${\bf{w}}_{TX} \in \mathbb{C}^{{P_t}{Q_t} \times 1}$ and ${\bf{w}}_{RX} \in \mathbb{C}^{{P_r}{Q_r} \times 1}$ are the JCS transmit and communication receive beamforming (BF) vectors, respectively; $\| {{\bf{w}}_{TX}} \|_2^2 = \| {{\bf{w}}_{RX}} \|_2^2 = 1$. In this paper, the low-complexity least-square (LS) method is used to generate ${\bf{w}}_{TX}$ and ${\bf{w}}_{RX}$ for BF. BS utilizes the known DL preambles, i.e., $d_{n,m} = \bar d_{n,m}$, to conduct beam alignment with MUE. When beam alignment is completed, ${\bf{w}}_{TX} = {c_1}{[ {{{\bf{a}}^T}({\bf{\tilde p}}_{TX,0})} ]^\dag }$ and ${\bf{w}}_{RX} = {c_2}{[ {{\bf{a}}({\bf{\tilde p}}_{RX,0})} ]^\dag }$, where ${c_1}$ and ${c_2}$ are both arbitrary complex values with modulus 1, ${\left[  \cdot  \right]^\dag }$ is the pseudo-inverse operation, ${\bf{\tilde p}}_{TX,0} \approx {\bf{p}}_{TX,0}$, and ${\bf{\tilde p}}_{RX,0} \approx {\bf{p}}^{C}_{RX,0} $. Simultaneously, the unknown communication CSI, $( {{\bf{w}}_{RX}} )^H{\bf{H}}_{C,n,m}{\bf{w}}_{TX}$, can be estimated by processing the received preambles.

Moreover, $n^{X}_{C,n,m} = n_{C,n,m} + \xi _{C,n,m}$ is the sum of noise and interference, $n_{C,n,m} = {\left( {{\bf{w}}_{RX}} \right)^H}{\bf{n}}_{C,n,m}$ and $\xi _{C,n,m} = {\left( {{\bf{w}}_{RX}} \right)^H}{\bf{x}}_{C,n,m}$ are transformed noise and interference, the dimensions of ${\bf{n}}_{C,n,m}$ and ${\bf{x}}_{C,n,m}$ are both ${P_r}{Q_r} \times 1$, ${\bf{n}}_{C,n,m}$ is Gaussian noise vectors with each element following ${\cal C}{\cal N}(0,\sigma _N^2)$, and ${\bf{x}}_{C,n,m}$ is the reflected interference signals from other network devices. 
We assume there are ${N_{ic}}$ ISs, and the reflected fading for each IS follows a Gaussian distribution. Since the superimposed one of multiple random OFDM signals is noise-like, the $p$th element of ${\bf{x}}_{C,n,m}$ can be given as ${\left[ {{\bf{x}}_{C,n,m}} \right]_p} = \sum\limits_{i = 0}^{{N_{ic}} - 1} {\sqrt {{P_{i,c}}} } \beta _{i,p}^I$, where ${P_{i,c}}$ is the power of incident signal from the $i$th IS, and $\beta _{i,p}^I \sim \mathcal{CN}(0,1)$. Let ${P_{IC}}{\rm{ = }}\sum\limits_{i = 0}^{{N_{ic}} - 1} {{P_{i,c}}}$. The interference to noise power ratio (INR) is $\gamma _C^{IN} = \frac{{{P_{IC}}}}{{\sigma _N^2}}$. Further, we define the communication SINR (C-SINR) as
\begin{equation}\label{equ:gamma_c}
	{\gamma _{C,n,m}} = \frac{{P_t\left\| {{h_{C,n,m}}} \right\|_2^2}}{{{P_{IC}} + \sigma _N^2}},
\end{equation}
where ${h_{C,n,m}} = {\left( {{\bf{w}}_{RX}} \right)^H}{\bf{H}}_{C,n,m}{\bf{w}}_{TX}$ is the gain of DL communication signal at each antenna element.
}

{\color{blue} 
\subsubsection{DL Echo Sensing Received Signal} \label{sec:DL Echo Sensing Received Signal}
The echo signal that BS receives for the $m$th OFDM symbol at the $n$th subcarrier is given by
\begin{equation}\label{equ:sensing_signal}
		{\bf{y}}_{S,n,m} = \sqrt {P_t} d_{n,m}{\bf{H}}_{S,n,m}{\bf{w}}_{TX} + {\bf{n}}_{S,n,m}^{X} = \sqrt {P_t} d_{n,m}\sum\limits_{l = 0}^{L - 1} {\left[\!\! \begin{array}{l}
				( {{\alpha _{S,n,m,l}}} )\chi _{TX,l}
				 {\bf{a}}( {{\bf{p}}_{RX,l}^{S}} )
			\end{array} \!\!\right]}  \!+\! {\bf{n}}_{S,n,m}^{X},
\end{equation}
where $\chi _{TX,l} = {{\bf{a}}^T}( {{\bf{p}}_{TX,l}} ){\bf{w}}_{TX}$ represents the gain of the DL JCS transmit BF, ${\bf{n}}_{S,n,m}^{X} = {\bf{n}}_{S,n,m} + {\bf{x}}_{S,n,m}$ is the sum of noise and interference, ${\bf{n}}_{S,n,m}$ is the Gaussian noise vector with each element following ${\cal C}{\cal N}(0,\sigma _N^2)$, ${\bf{x}}_{S,n,m}$ is the superimposed interference vector for ${N_{is}}$ reflected ISs, and the dimensions of ${\bf{n}}_{S,n,m}$ and ${\bf{x}}_{S,n,m}$ are ${P_t}{Q_t} \times 1$. Similar to ${\bf{x}}_{C,n,m}$, the $p$th element of ${\bf{x}}_{S,n,m}$ can be given as ${\left[ {{\bf{x}}_{S,n,m}} \right]_p} = \sum\limits_{i = 0}^{{N_{is}} - 1} {\sqrt {{P_{i,s}}} } \beta _{i,p}^I$, where ${P_{i,s}}$ is the incident power of the $i$th IS, and $\beta _{i,p}^I \sim \mathcal{CN}(0,1)$. The aggregate power of each element of ${\bf{x}}_{S,n,m}$ is ${P_{IS}}{\rm{ = }}\sum\limits_{i = 0}^{{N_{ic}} - 1} {{P_{i,s}}}$. The sensing INR is defined as $\gamma _S^{IN} = {{{P_{IS}}} \mathord{/
		{\vphantom {{{P_{IS}}} {\sigma _N^2}}} 
		\kern-\nulldelimiterspace} {\sigma _N^2}}$. Further, the sensing SINR (S-SINR) is defined as 
\begin{equation}\label{equ:gamma_s}
	{\gamma _{S,n,m}} = \frac{{P_t\left\| {{h_{S,n,m,l}}} \right\|_2^2}}{{{P_{IS}} + \sigma _N^2}},
\end{equation}
where ${h_{S,n,m,l}} = {\alpha _{S,n,m,l}}\chi _{TX,l}$ is the gain of DL echo sensing signal at each antenna element.

By defining $s_{n,m,l} = \sqrt {P_t} d_{n,m}{\alpha _{S,n,m,l}}\chi _{TX,l}$ and ${\bf{s}}_{n,m} = { {[ {s_{n,m,l}} ]} |_{l = 0,1,...,L - 1}}$, \eqref{equ:sensing_signal} can be expressed in the matrix form as
\begin{equation}\label{equ:y_s_n_m}
	{\bf{y}}_{S,n,m} = {{\bf{A}}_{S,RX}}{\bf{s}}_{n,m} + {\bf{n}}_{S,n,m}^{X},
\end{equation}
where ${{\bf{A}}_{S,RX}} = { {[ {{\bf{a}}( {{\bf{p}}_{RX,l}^{S}} )} ]} |_{l = 0,1,...,L - 1}}$ is the steering matrix stacked by steering vectors of $L$ echoes, ${{\bf{A}}_{S,RX}} \in \mathbb{C}^{{P_t}{Q_t} \times L}$, and ${\bf{s}}_{n,m}\in \mathbb{C}^{L \times 1}$.
By stacking all the $M_s$ OFDM symbols with $N_c$ subcarriers, we have
\begin{equation}\label{equ:y_s_D}
	{\bf{Y}}_S = {{\bf{A}}_{S,RX}}{{\bf{S}}} + {\bf{N}}_{t}^{X},
\end{equation}
where ${{\bf{S}}} = { {[ {{\bf{s}}_{n,m}} ]} |_{(n,m) \in [0, \cdots ,N_c] \times [0, \cdots ,M_s]}} \in \mathbb{C}^{L \times N_cM_s}$, and ${\bf{Y}}_S \in \mathbb{C}^{{P_t}{Q_t} \times {N_c}{M_s}}$.

}

\section{DL JCS Signal Processing}\label{sec:JCS Downlink signal processing}
In this section, we demonstrate the signal processing for DL JCS sensing and communication, which is shown in Fig.~\ref{fig: DL_JCS_signal_processing}. We first present the sensing signal processing scheme, and then elaborate on the JCAS CSI enhancement method.

%

\setcounter{equation}{13}
\subsection{JCS Sensing Signal Processing}
In this subsection, we first present the conventional MUSIC method for estimating the 2D AoAs, and then introduce the novel MUSIC-based range and Doppler estimation method.

\subsubsection{JCS MUSIC 2D Angle Detection}
First, the correlation matrix of ${\bf{Y}}_S$ is obtained as
\begin{equation}\label{equ:R_x_D}
	{\bf{R}}_{\bf{X}}{\rm{ = }}\frac{1}{{M_sN_c}} {{\bf{Y}}_S{{[ {{\bf{Y}}_S} ]}^H}} .
\end{equation}
By applying eigenvalue decomposition to ${\bf{R}}_{\bf{X}}$, we have
\begin{equation}\label{equ:svd_R_x_D}
	\left[ {\bf{U}}_x,{\bf{\Sigma }}_x \right] = {\rm eig}\left( {{\bf{R}}_{\bf{X}}} \right),
\end{equation}
where ${\bf{\Sigma }}_x$ is the real-value eigenvalue diagonal matrix in descending order, and ${\bf{U}}_x$ is the orthogonal eigen matrix. Calculate the average of eigenvalues and denote it as $m_x$. {\color{blue} Let ${\alpha _t}$ be a preset threshold, which is determined as elaborated in \textbf{Appendix~\ref{Appendix_alpha_t}}.} Then, the number of echo paths is determined as the number of eigenvalues no smaller than ${\alpha _t}m_x$, denoted by $N_x$. 

Construct ${\bf{U}}_N = {\bf{U}}_x\left( {:,N_x + 1:{P_t}{Q_t}} \right)$\footnote{${\bf{U}}_x\left( {:,N_x + 1:{P_t}{Q_t}} \right)$ means the slice matrix of  $(N_x + 1)$th to the ${P_t}{Q_t}$th columns of the matrix.} as the noise subspace basis. We then use it to obtain the spatial angular spectrum function as~\cite{HAARDT2014651}
\begin{equation}\label{equ:spectrum function}
	f_a\left( {{\bf{p}};{\bf{U}}_N} \right) = {{\bf{a}}^H}\left( {{\bf{p}}} \right) {{\bf{U}}_N{{\left( {{\bf{U}}_N} \right)}^H}} {\bf{a}}\left( {{\bf{p}}} \right),
\end{equation}
where ${\bf{p}} = \left( {\varphi ,\theta } \right)$ is the 2D angle, and ${\bf{a}}\left( \bf{p} \right)$ is given in \eqref{equ:steeringVec}. The spatial spectrum is represented as \cite{HAARDT2014651}
\begin{equation}\label{equ:spatial_spectrum}
	S_a\left( {{\bf{p}};{\bf{U}}_N} \right) = {[ {{{\bf{a}}^H}\left( {\bf{p}} \right){\bf{U}}_N{{( {{\bf{U}}_N} )}^H}{\bf{a}}\left( {\bf{p}} \right)} ]^{ - 1}}.
\end{equation}
The maximum points of $S_a\left( {{\bf{p}};{\bf{U}}_N} \right)$, i.e., the minimum points of $f_a\left( {{\bf{p}};{\bf{U}}_N} \right)$ are the estimated AoAs~\cite{Lifu1993}. We first find $N_x$ local maximum points of $S_a\left( {{\bf{p}};{\bf{U}}_N} \right)$ using a grid searching method with relatively large granularity, then we use the Newton descent method to identify the accurate minimum point of $f_a\left( {{\bf{p}};{\bf{U}}_N} \right)$ by inputting the above local maximum points as initial points for iteration.

\subsubsection{JCS Range and Doppler Detection}
{\color{blue}
After the AoAs are obtained, through BF at the AoA of interest, the filtered received signal at the $n$th subcarrier of the $m$th OFDM symbol can be expressed as
\begin{equation}\label{equ:beamforming_receive}
		\bar y_{S,n,m,k} = {( {{\bf{w}}_{RX,S,k}} )^H}{\bf{y}}_{S,n,m} = \sqrt {P_t} d_{n,m}\sum\limits_{l = 0}^{L - 1} {[ {{\alpha _{S,n,m,l}}\chi _{TX,l}\varpi _{RX,l,k}} ]}  + w_{t,n,m,k},
\end{equation}
where $w_{t,n,m,k}{\rm{ = }}{( {{\bf{w}}_{RX,S,k}} )^H}{\bf{n}}_{S,n,m}^{X}$ is the transformed noise and interference with zero mean and variance ${\sigma _W}^2$, $\varpi _{RX,l,k} = {( {{\bf{w}}_{RX,S,k}} )^H}{\bf{a}}( {{\bf{p}}_{RX,l}^{S}} )$ is the receive BF gain, and ${{\bf{w}}_{RX,S,k}}$ is the receive BF vector for the $k$th AoA, $k \in [ {0,1,...,N_x - 1} ]$. Note that $\varpi _{RX,k,k}$ is typically larger than $\varpi _{RX,l,k}$ ($l \ne k$) due to the narrow beam feature of mmWave.
}

By substituting \eqref{equ:alpha_S} into \eqref{equ:beamforming_receive}, we obtain \eqref{equ:y_n_m_k}.
\begin{equation}\label{equ:y_n_m_k}
	\begin{array}{l}
		\bar y_{S,n,m,k} = \sqrt {P_t} d_{n,m}{b_{S,k}}\varpi _{RX,k,k}\chi _{TX,k}{e^{j2\pi {f_{s,k,1}}m{T}}}{e^{ - j2\pi n\Delta {f}\left( {\frac{{{r_k}}}{c}} \right)}} + \\
		\sum\limits_{l = 0,l \ne k}^{L - 1} {\left[ {\sqrt {P_t} d_{n,m}{b_{S,l}}\varpi _{RX,l,k}\chi _{TX,l}{e^{j2\pi {f_{s,l,1}}m{T}}}{e^{ - j2\pi n\Delta {f}\left( {\frac{{{r_l}}}{c}} \right)}}} \right]}  + w_{t,n,m,k}.
	\end{array}
\end{equation}
In \eqref{equ:y_n_m_k}, there are independent complex exponential functions for range and Doppler, i.e., $e^{ - j2\pi n\Delta {f}\left( {\frac{{{r_l}}}{c}} \right)}$ and $e^{j2\pi {f_{s,l,1}}m{T}}$, respectively. 
\setcounter{equation}{19}
Here, we define the range and Doppler steering vectors as 
\begin{equation}\label{equ:range_steering}
	{{\bf{a}}_r}\left( r \right) = { {[ {{e^{ - j2\pi n\Delta {f}\frac{r}{c}}}} ]} |_{n = 0,1,...,{N_c} - 1}},
\end{equation}
\begin{equation}\label{equ:doppler_steering}
	{{\bf{a}}_f}\left( f \right) = { {[ {{e^{j2\pi m{T}f}}} ]} |_{m = 0,1,...,{M_s} - 1}},
\end{equation}
respectively. The range and Doppler steering matrices are defined as
\begin{equation}\label{equ:range_steering_matrix}
	{{\bf{A}}_{\bf{r}}} = { {[ {{{\bf{a}}_r}\left( {{r_l}} \right)} ]} |_{l = 0,1,...,L - 1}},
\end{equation}
\begin{equation}\label{equ:Doppler_steering_matrix}
	{{\bf{A}}_{\bf{f}}} = { {[ {{{\bf{a}}_f}\left( {{f_{s,l,1}}} \right)} ]} |_{l = 0,1,...,L - 1}},
\end{equation}
where ${{\bf{A}}_{\bf{r}}} \in \mathbb{C}^{N_c \times L}$, and ${{\bf{A}}_{\bf{f}}} \in \mathbb{C}^{M_s \times L}$.

Stack $\bar y_{S,n,m,k}$ into a matrix ${\bf{\bar Y}}_S$ where ${\left[ {{\bf{\bar Y}}_S} \right]_{n,m}} = \bar y_{S,n,m,k}$, then erase the communication symbol matrix ${\bf{D}}_s$ where ${\left[ {{\bf{D}}_s} \right]_{n,m}} = d_{n,m}$. From ${\bf{\bar Y}}_S$, we obtain
\begin{equation}\label{equ:Y_SSU}
	{\bf{\bar H}}_S = \frac{{{\bf{\bar Y}}_S}}{{{\bf{D}}_s}},
\end{equation}
where the division is element-wise, and ${\bf{\bar H}}_S \in \mathbb{C}^{N_c \times M_s}$.

According to \eqref{equ:y_n_m_k}, ${\bf{\bar H}}_S$ can be expressed by ${{\bf{A}}_{\bf{r}}}$ as 
\begin{equation}\label{equ:Y_SSU_As}
	{\bf{\bar H}}_S = {{\bf{A}}_{\bf{r}}}{\bf{S}}_{r,s} + {\bf{W}}_{tr},
\end{equation}
where ${\bf{S}}_{r,s} = { {[ {{\bf{s}}_{r,m}} ]} |_{m = 0,1,...,{M_s} - 1}} \in \mathbb{C}^{L \times M_s}$, ${\bf{s}}_{r,m} = { {[ {\sqrt {P_t} {b_{S,l}}\varpi _{TX,l,k}\chi _{TX,l}} {e^{j2\pi m{T} {f_{s,l,1}}}} ]} |_{l = 0,1,...,L - 1}}$, and $\left[ {\bf{W}}_{tr} \right]_{n,m} = w_{t,n,m,k}$. 

On the other hand, the transpose of ${\bf{\bar H}}_S$, i.e., ${\left( {\bf{\bar H}}_S \right)^T}$, can be presented by ${{\bf{A}}_{\bf{f}}}$ as
\begin{equation}\label{equ:Y_SSU_TAs}
	\left( {\bf{\bar H}}_S \right)^T = {{\bf{A}}_{\bf{f}}}{\bf{S}}_{f,s} + {\bf{W}}_{tf},
\end{equation}
where ${\bf{S}}_{f,s} = { {[ {{\bf{s}}_{f,n}} ]} |_{n = 0,1,...,N_c - 1}} \in \mathbb{C}^{L \times N_c}$, ${\bf{s}}_{f,n} \!\!=\!\! { {[ {\sqrt {P_t} {b_{S,l}}\varpi _{TX,l,k}\chi _{TX,l}}{e^{ - j2\pi n\Delta {f} {\frac{{{r_l}}}{c}} }} ]} |_{l = 0,1,...,L - 1}}$, and ${\bf{W}}_{tf} = {[ {{\bf{W}}_{tr}} ]^T}$.

{\color{blue} 
The range and Doppler can be estimated via the autocorrelation of ${\bf{\bar H}}_S$ and ${\left( {{\bf{\bar H}}_S} \right)^T}$, which are given by
\begin{equation}\label{equ:R_x_tau}
		{{\bf{R}}_{{{X}},r}} = \frac{1}{{{M_s}}}{\bf{\bar H}}_S{( {{\bf{\bar H}}_S} )^H},
		{{\bf{R}}_{{{X}},f}} = \frac{1}{{{N_c}}}{( {{\bf{\bar H}}_S} )^T}{( {{\bf{\bar H}}_S} )^*},
\end{equation}
respectively.
Denote the noise subspaces of ${{\bf{R}}_{{{X}},r}}$ and ${{\bf{R}}_{{{X}},f}}$ as ${{\bf{U}}_{x,rN}}$ and ${{\bf{U}}_{x,fN}}$, respectively.

\begin{Theo} \label{Theo:1}
	{\rm 
		The minimum of ${\| {{\bf{U}}{{_{x,r N}}^H}{{\bf{a}}_r}\left( r \right)} \|_2^2}$, denoted by ${r_{s,l}}$, is linked to the range via ${r_{s,l}} = 2{d_{l,1}}$. The minimum of ${\| {{\bf{U}}{{_{x,fN}}^H}{{\bf{a}}_f}\left( f \right)} \|_2^2}$ corresponds to the Doppler value, ${f_{s,l,1}}$.
		\begin{proof}
			The proof is presented in Appendix \ref{Theo:A}. 
		\end{proof}
	}
\end{Theo}

By applying eigenvalue decomposition to ${\bf{R}}_{{{X}},r }$ and ${\bf{R}}_{{{X}},f}$, we have
\begin{equation}\label{equ:ED_R_x_tau}
	\begin{aligned}
		\left[ {{\bf{U}}_{x,r },{\bf{\Sigma }}_{x,r }} \right] = {\rm eig}\left( {{\bf{R}}_{{{X}},r }} \right),\
		\left[ {{\bf{U}}_{x,f},{\bf{\Sigma }}_{x,f}} \right] = {\rm eig}\left( {{\bf{R}}_{{{X}},f}} \right),
	\end{aligned}
\end{equation}
where ${\bf{\Sigma }}_{x,r }$ and ${\bf{\Sigma }}_{x,f}$ are the real-value diagonal matrices of eigenvalues in the descending order, and ${\bf{U}}_{x,r }$ and ${\bf{U}}_{x,f}$ are the corresponding eigenvector matrices. 

We use $m_{x,r }$ to denote the mean value of ${\bf{\Sigma }}_{x,r }$, and then set the threshold ${\alpha _{t,r}}$ using the method in \textbf{Appendix}~\ref{Appendix_alpha_t} by replacing ${\bf{\Sigma }}_x$ with ${\bf{\Sigma }}_{x,r }$. The number of targets in the AoA of interest, $N_{x,r }$, is then determined as the number of eigenvalues no smaller than ${\alpha _{t,r }}m_{x,r }$. Then, the noise subspace basis for range estimation is derived as ${\bf{U}}_{x,r N} = {\bf{U}}_{x,r }( {:,N_{x,r } + 1:{N_c}} )$. Since the number of targets is the same for both Doppler and range estimation, the noise subspace basis for the Doppler estimation can be derived as ${\bf{U}}_{x,fN} = {\bf{U}}_{x,f}( {:,N_{x,r} + 1:M_s} )$. 

We use ${\bf{U}}_{x,r N}$ and ${\bf{U}}_{x,fN}$ to derive the range and Doppler spectrum functions as 
\begin{equation}\label{equ:fr}
	\begin{aligned}
		f_r( r;{\bf{U}}_{x,r N} ) = {{\bf{a}}_r}{( r )^H}{\bf{U}}_{x,r N}{( {{\bf{U}}_{x,r N}} )^H}{{\bf{a}}_r}( r ),\
		f_f( f;{\bf{U}}_{x,fN} ) = {{\bf{a}}_f}{( f )^H}{\bf{U}}_{x,fN}{( {{\bf{U}}_{x,fN}} )^H}{{\bf{a}}_f}( f ),
	\end{aligned}	
\end{equation}
respectively. The range and Doppler spectra can be given by
\begin{equation}\label{equ:Sr}
	\begin{aligned}
		{S_r}( {r;{\bf{U}}_{x,rN}} ) = {[ {{{\bf{a}}_r}{{( r )}^H}{\bf{U}}_{x,rN}{{( {{{\bf{U}}_{x,rN}}} )}^H}{{\bf{a}}_r}( r )} ]^{ - 1}},\\
		{S_f}( {f;{\bf{U}}_{x,fN}} ) = {[ {{{\bf{a}}_f}{{( f )}^H}{\bf{U}}_{x,fN}{{( {{\bf{U}}_{x,fN}} )}^H}{{\bf{a}}_f}( f )} ]^{ - 1}},
	\end{aligned}
\end{equation}
respectively.

The maximum points of $S_r( r;{\bf{U}}_{x,r N} )$ and $S_f( f;{\bf{U}}_{x,fN} )$, i.e., the minimum points of $f_r( r;{\bf{U}}_{x,r N} )$ and $f_f( f;{\bf{U}}_{x,fN} )$, are the range and Doppler estimation values, denoted by ${\hat r_{s,l}}$ and $\hat f_{s,l}$, respectively. The distance, $d_{l,1}$, and radial velocity, $v_{r,0,1}$ , between BS and the target are given by ${\hat d_{l,1}} = \frac{{{{\hat r}_{s,l}}}}{2}$ and ${\hat v_{r,0,1}} = \frac{{\lambda {{\hat f}_{s,l}}}}{2}$.}

The minimum of $f_r( r )$ and $f_f( f )$ can be identified using a two-step Newton descent method.
We first find the local maximum points of $S_r( r )$ and $S_f( f )$ with large-granularity grid searching. Then, we use the Newton descent method to find the accurate minimum points of $f_r( r )$ and $f_f( f )$ using the above local maximum points as the initial points. The iterative expression for the Newton descent method is derived as follows. 

By applying the Taylor series decomposition to $f_r( r )$ and $f_f( f )$, and taking their first order derivative over $r$ and $f$, respectively, we obtain 
\begin{equation}\label{equ:Taylor_fr}
\frac{{\partial f_r\left( {r} \right)}}{{\partial r}} \buildrel\textstyle.\over= \frac{{\partial f_r( {{r_0}} )}}{{\partial r}} + \frac{{{\partial ^2}[ {f_r( {{r_0}} )} ]}}{{{\partial ^2}r}}( {r - {r_0}} ),
\end{equation}
and 
\begin{equation}\label{equ:Taylor_ff}
	\frac{{\partial f_f( {f} )}}{{\partial f}} \buildrel\textstyle.\over= \frac{{\partial f_f( {{f_0}} )}}{{\partial f}} + \frac{{{\partial ^2}[ {f_f( {{f_0}} )} ]}}{{{\partial ^2}f}}( {f - {f_0}} ).
\end{equation}
By setting the above first-order derivative to be 0, the iterative descent expression for range and Doppler estimation can be given by
\begin{equation}\label{equ:descent_r}
	{r^{( k )}} = {r^{( {k - 1} )}} - {\left[ {\frac{{{\partial ^2}f_r( {{r^{( {k - 1} )}}} )}}{{{\partial ^2}r}}} \right]^{ - 1}}\frac{{\partial f_r( {{r^{( {k - 1} )}}} )}}{{\partial r}},
\end{equation}
\begin{equation}\label{equ:descent_f}
	{f^{( k )}} = {f^{( {k - 1} )}} - {\left[ {\frac{{{\partial ^2}f_f( {{f^{( {k - 1} )}}} )}}{{{\partial ^2}f}}} \right]^{ - 1}}\frac{{\partial f_f( {{f^{( {k - 1} )}}} )}}{{\partial f}},
\end{equation}
respectively.
From \eqref{equ:fr}, the first-order and second-order derivatives of ${f_f\left( f \right)}$ and ${f_r\left( r \right)}$ are expressed as
\begin{equation}\label{equ:first_derivative_f_r}
	\frac{{\partial f_r( r )}}{{\partial r}} {\rm{ = }}2{\mathop{\rm Re}\nolimits} \{ {{{\bf{a}}^{( 1 )}_r}{{( r )}^H}{\bf{U}}_{x,r N}{{( {{\bf{U}}_{x,r N}} )}^H}{{\bf{a}}_r}( r )} \},
\end{equation}
\begin{equation}\label{equ:first_derivative_f_f}
	\frac{{\partial f_f( f )}}{{\partial f}}{\rm{ = }}2{\mathop{\rm Re}\nolimits} \{ {{\bf{a}}_f^{( 1 )}{{( f )}^H}{\bf{U}}_{x,fN}{{( {{\bf{U}}_{x,fN}} )}^H}{{\bf{a}}_f}( f )} \},
\end{equation}
\begin{equation}\label{equ:second_derivative_f_r}
	\frac{{{\partial ^2}f_r( r )}}{{{\partial ^2}r}} \!=\! 2{\mathop{\rm Re}\nolimits} \!\left\{\!\!\! \begin{array}{l}
		{{\bf{a}}_r^{( 2 )}}{( r )^H}{\bf{U}}_{x,r N}{( {{\bf{U}}_{x,r N}} )^H}{{\bf{a}}_r}( r )
		+ {{\bf{a}}_r^{( 1 )}}{( r )^H}{\bf{U}}_{x,r N}{( {{\bf{U}}_{x,r N}} )^H}{{\bf{a}}^{( 1 )}_r}( r )
	\end{array} \!\!\! \right\},
\end{equation}
\begin{equation}\label{equ:second_derivative_f_f}
	\frac{{{\partial ^2}f_f\left( f \right)}}{{{\partial ^2}f}}\! = \!2{\mathop{\rm Re}\nolimits} \!\left\{
		{\bf{a}}_f^{( 2 )}{( f )^H}{\bf{U}}_{x,fN}{( {{\bf{U}}_{x,fN}} )^H}{{\bf{a}}_f}( f )
		+ {\bf{a}}_f^{( 1 )}{( f )^H}{\bf{U}}_{x,fN}{( {{\bf{U}}_{x,fN}} )^H}{\bf{a}}_f^{( 1 )}( f ) \right\},
\end{equation}
where ${\bf{a}}_r^{( 1 )}{{( r )}}$, ${\bf{a}}_f^{( 1 )}{{( f )}}$, ${\bf{a}}_r^{( 2 )}{( r )}$, and ${\bf{a}}_f^{( 2 )}( f )$ are the first-order and second-order derivatives of ${{\bf{a}}_r}( r )$ and ${{\bf{a}}_f}( f )$, respectively. From \eqref{equ:range_steering} and \eqref{equ:doppler_steering}, these expressions are presented as
\begin{equation}\label{equ:first_derivative_a}
	\begin{array}{c}
		{\bf{a}}_r^{( 1 )}( r ) = { {[ {( { - j2\pi n\frac{{\Delta {f}}}{c}} ){e^{ - j2\pi n\Delta {f}\frac{r}{c}}}} ]} |_{n = 0,1,...,{N_c} - 1}},
		{\bf{a}}_f^{( 1 )}( f ) = { {[ {( {j2\pi m{T}} ){e^{j2\pi m{T}f}}} ]} |_{m = 0,1,...,{M_s} - 1}},\\
		{\bf{a}}_r^{( 2 )}( r ) = { {[\! {{{( { - j2\pi n\frac{{\Delta f}}{c}} )}^2}{e^{ - j2\pi n\Delta {f}\frac{r}{c}}}} ]}\! |_{n = 0,1,...,{N_c} - 1}},
		{\bf{a}}_f^{( 2 )}( f ) = { {[ {{{( {j2\pi m{T}} )}^2}{e^{j2\pi m{T}f}}} ]}\! |_{m = 0,1,...,{M_s} - 1}}.
	\end{array}
\end{equation}

{\color{blue}
\subsection{JCS Communication Signal Processing} \label{sec:JCS_comm}
By substituting \eqref{equ:general_JCS_channel} into \eqref{equ:y_DC}, and taking into consideration that ${\bf{w}}_{TX}$ and ${\bf{w}}_{RX}$ in \eqref{equ:y_DC} generate beams pointed at the AoD and AoA of the LoS communication path, respectively, we obtain the communication received signal as 
\begin{equation}\label{equ:y_Cnm}
	y_{C,n,m} = \sqrt {P_t} d_{n,m}h_{C,n,m} + n^{X}_{C,n,m},
\end{equation}
where $h_{C,n,m} = {b_{C,0}}\varpi _{RX,0}\chi _{TX,0}{e^{j2\pi mT_s{f_{c,d,0}}}}{e^{ - j2\pi n\Delta {f}{\tau _{c,0}}}}$ is the real communication channel response, and $\varpi _{RX,0} = {\left( {{\bf{w}}_{RX}} \right)^H} {\bf{a}}( {{\bf{p}}^{C}_{RX,0}} )$ and $\chi _{TX,0} = {{\bf{a}}^T}({\bf{p}}_{TX,0}){\bf{w}}_{TX}$ are the BF transmitting and receiving gains. In the CSI estimation, $d_{n,m} = \bar d_{n,m}$, and we denote $y_{C,n,m} = \bar y_{C,n,m}$ as the received signal. The CSI estimated with the LS method is expressed as \cite{2010MIMO}
\begin{equation}\label{equ:h_CD_nm_hat}
	\hat h_{C,n,m} = \frac{{\bar y_{C,n,m}}}{{\sqrt {P_t} \bar d_{n,m}}} = h_{C,n,m} + w_{C,n,m},
\end{equation}
where $w_{C,n,m} = \frac{n^{X}_{C,n,m}}{{\sqrt {P_t} \bar d_{n,m}}}$ is the transformed noise plus interference and follows $\mathcal{CN}(0, \sigma _p^2)$, $\sigma _p^2 = ({P_{IC}} + \sigma _N^2)/P_t$. The estimated communication response matrix at $M_s$ OFDM symbols is denoted by ${\bf{\hat H}}_C$, where ${[ {{\bf{\hat H}}_C} ]_{n,m}} = \hat h_{C,n,m}$. The method for estimating $\sigma _p^2$ based on ${\bf{\hat H}}_C$ is presented in \textbf{Appendix}~\ref{Appendix_sigma_p}.

The conventional communication uses ${\bf{\hat H}}_C$ to demodulate the communication data. On the other hand, ${f_{c,d,0}}$ and ${\tau _{c,0}}$ can be estimated by JCS as ${\hat f_{c,d,0}} = {{{\hat f_{s,0}}} \mathord{\left/
		{\vphantom {{{f_{s,0}}} 2}} \right.
		\kern-\nulldelimiterspace} 2}$ and ${{\hat \tau }_{c,0}} = {{{\hat r_{s,0}}} \mathord{\left/
		{\vphantom {{{r_{s,0}}} {(2c)}}} \right.
		\kern-\nulldelimiterspace} {(2c)}}$,
respectively. Based on the prior information obtained by JCS sensing, we propose a Kalman filter-based JCS CSI enhancement method to improve CSI by leveraging the sensing estimation results of JCS. 

For the $m$th OFDM symbol, $\hat h_{C,n,m}$ can be regarded as the observation of $h_{C,n,m}$ as given in~\eqref{equ:h_CD_nm_hat}. Since ${b_{C,0}}$ is unchanged for the same OFDM symbol. The state transfer of $h_{C,n,m}$ is given by 
\begin{equation}\label{equ:state_trasfer_h_dc}
	h_{C,n + 1,m} = {e^{ - j2\pi \Delta {f}\left( {{\tau _{c,0}}} \right)}}h_{C,n,m},
\end{equation}

The Kalman filter algorithm that utilizes $\hat \Phi  = { {[ {\hat h_{C,n,m}} ]} |_{n = 0, \cdots ,N_c - 1}}$ to recursively derive the estimation of $\Phi  = { {[ {h_{C,n,m}} ]} |_{n = 0, \cdots ,N_c - 1}}$ is presented in \textbf{Algorithm~\ref{DL_Kalman_CSI}}, with the details of the Kalman Filter algorithm referenced to~\cite{2017Kalman}. Note that we obtain $h_{C,n,m} = {e^{ - j2\pi n\Delta {f} {{\tau _{c,0}}} }}h_{C,0,m}$ from \eqref{equ:state_trasfer_h_dc}, based on which we can further estimate the initial observation variance as
\begin{equation}\label{equ:init_obs_pw}
	{p_{w,0}} \!\!=\!\!\!\! {{\sum\limits_{n = 1}^{N_c - 1} {\| {{e^{j2\pi n\Delta {f}\left( {{{\hat \tau }_{c,0}}} \right)}}\hat h_{C,n,m} \!-\! \hat h_{C,0,m}} \|_2^2} } \mathord{/
			{\vphantom {{\sum\limits_{n = 1}^{N_c - 1} {\left\| {{e^{j2\pi n\Delta {f}\left( {{{\hat \tau }_{c,0}}} \right)}}\hat h_{C,n,m} - \hat h_{C,0,m}} \right\|_2^2} } {(N_c - 1)}}} 
			\kern-\nulldelimiterspace} {(N_c - 1)}},
\end{equation}

\begin{algorithm}[!t]
	\caption{JCS CSI Enhancement method}
	\label{DL_Kalman_CSI}
	\KwIn{The observation variance $\sigma _p^2$; The variance of initial estimation ${p_{w,0}}$; The initial observation $\hat h_{C,0,m}$; The transfer factor $A = {e^{ - j2\pi \Delta {f}{{\hat \tau }_{c,0}}}}$; The observation sequence $\hat \Phi$.
	}
	\KwOut{Filtered sequence ${ {[ {\bar h_{C,n,m}} ]} |_{n = 0, \cdots ,N_c - 1}}$.}
	\textbf{Step} 1: $\bar h_{C,0,m} = \hat h_{C,0,m}$.
	
	\textbf{Step} 2: \For{$n$ = {\rm 1} to $N_c - 1$} {
		$\hat h_{n,m}^ -  = A\bar h_{C,n-1,m}$\;
		$p_{w,n}^ -  = A{p_{w,n - 1}}{A^*}$\;
		${K_k} = {( {p_{w,n}^ - } )^*}{( {p_{w,n}^ -  + \sigma _p^2} )^{ - 1}}$\;
		$\bar h_{C,n,m} = \hat h_{n,m}^ -  + ( {\hat h_{C,n,m} - \hat h_{n,m}^ - } ) {K_k}$\;
		${p_{w,n}} = ( {1 - {K_k}} )p_{w,n}^ - $\;
	}	
	\Return ${ {[ {\bar h_{C,n,m}} ]} |_{n = 0, \cdots ,N_c - 1}}$.
\end{algorithm}

After ${ {[ {\bar h_{C,n,m}} ]} |_{n = 0, \cdots ,N_c - 1}}$ for $m = 0,...,M_s - 1$ are all derived via \textbf{Algorithm~\ref{DL_Kalman_CSI}}, we can form the enhanced CSI matrix ${\bf{\bar H}}_C$, where ${[{\bf{\bar H}}_C]_{n,m}} = \bar h_{C,n,m}$, to demodulate the data symbols. First, $y_{C,n,m}$ given in \eqref{equ:y_DC} is equalized as ${\hat r_{C,n,m}} = {{y_{C,n,m}} \mathord{/
		{\vphantom {{y_{C,n,m}} {(\sqrt {P_t} \bar h_{C,n,m})}}} 
		\kern-\nulldelimiterspace} {(\sqrt {P_t} \bar h_{C,n,m})}}$, then we use the maximum likelihood (ML) method to estimate $d_{n,m}$ as ${\hat d_{n,m}} = \mathop {\arg \min }\limits_{d \in {\Theta _{QAM}}} \| {{{\hat r}_{C,n,m}} - d} \|_2^2$, where ${\Theta _{QAM}}$ is the constellation.

}

\section{Performance analysis of the JCS Processing}\label{sec:JCS performance}
In this section, the analytical MSE results of AoAs, range, Doppler, and location estimation of the proposed MUSIC-based JCS processing are derived using the perturbation method.

\subsection{Analysis of 2D Angle Detection MSE}
\setcounter{equation}{43}
From \eqref{equ:y_s_D}, the noise term ${\bf{N}}_t$ can be treated as the perturbation to the useful signal, which is expressed as 
\begin{equation}\label{equ:noised_signal}
	{\bf{Y}}_S = {\bf{Y}}_{S,R} + {\bf{N}}_{t}^{X},
\end{equation}
where ${\bf{Y}}_{S,R} = {{\bf{A}}_{S,RX}}{{\bf{S}}}$ is the useful signal.
The singular value decomposition of ${\bf{Y}}_{S,R}$ can be expressed as 
\begin{equation}\label{equ:svd_of_useful_signal}
	{\bf{Y}}_{S,R} \!=\! {\bf{U\Sigma }}{{\bf{V}}^H} \!=\! \left[ {{{\bf{U}}_s},\!{{\bf{U}}_0}} \right]\!\left[\! { {\begin{array}{*{20}{c}}
				{{{\bf{\Sigma }}_s}}&{\bf{0}}\\
				{\bf{0}}&{\bf{0}}
		\end{array}} \!} \right]\!\!\left[\! {\begin{array}{*{20}{c}}
			{{\bf{V}}_s^H}\\
			{{\bf{V}}_0^H}
	\end{array}} \!\right] \!\!=\!\!{{\bf{U}}_s}{{\bf{\Sigma }}_s}{\bf{V}}_s^H,
\end{equation}
where ${{\bf{U}}_0}$ is the noise subspace basis, and ${{\bf{U}}_0}^H{\bf{Y}}_{S,R} = {\bf{0}}$. Further, we have
${{\bf{U}}_0}^H{{\bf{A}}_{S,RX}} = {\bf{0}}$. 

With noise as perturbation, ${\bf{Y}}_S$ can be expressed as
\begin{equation}\label{equ:svd_of_YSD}
	\begin{aligned}
		{\bf{Y}}_S = \left[ {{{{\bf{\tilde U}}}_s},{{{\bf{\tilde U}}}_0}} \right]\left[ { {\begin{array}{*{20}{c}}
					{{{{\bf{\tilde \Sigma }}}_s}}&{\bf{0}}\\
					{\bf{0}}&{{{{\bf{\tilde \Sigma }}}_{\bf{0}}}}
			\end{array}}} \right]\left[ {\begin{array}{*{20}{c}}
				{{\bf{\tilde V}}_s^H}\\
				{{\bf{\tilde V}}_0^H}
		\end{array}} \right],
	\end{aligned}
\end{equation}
where ${{\bf{\tilde \Sigma }}_{\bf{0}}} = {\bf{\Delta }}{{\bf{\Sigma }}_{\bf{0}}}$, ${{\bf{\tilde \Sigma }}_s} = {{\bf{\Sigma }}_s} + {\bf{\Delta }}{{\bf{\Sigma }}_s}$, and ${{\bf{\tilde U}}_0} = {{\bf{U}}_0} + \Delta {{\bf{U}}_0}$. Here, ${{\bf{\tilde U}}_0}$ and ${{\bf{\tilde U}}_s}$ are both orthogonal unitary matrices, and ${{\bf{\tilde U}}_0}^H{\bf{Y}}_S{\rm{ = }}{\bf{\Delta }}{{\bf{\Sigma }}_{\bf{0}}}{\bf{\tilde V}}_0^H$. In the high SINR regime, solving the perturbation problem is equivalent to seeking the optimal $\Delta {\bf{U}}_0$ to minimize $\| {{{{\bf{\tilde U}}}_0}^H{\bf{Y}}_S} \|_2$ subject to the constraint ${{\bf{\tilde U}}_0}^H{{\bf{\tilde U}}_0} = {\bf{I}}$~\cite{Lifu1993}. By substituting \eqref{equ:noised_signal} into $\| {{{{\bf{\tilde U}}}_0}^H{\bf{Y}}_S} \|_2$, we have
\begin{equation}\label{equ:U_YS}
	\| {{{{\bf{\tilde U}}}_0}^H{\bf{Y}}_S} \|_2{\rm{ = }}\| {{{( {{{\bf{U}}_0} + {\bf{\Delta }}{{\bf{U}}_{{0}}}} )}^H}( {{\bf{Y}}_{S,R} + {\bf{N}}_t^{X}} )} \|_2.
\end{equation}
The second-order perturbation ${( {{\bf{\Delta }}{{\bf{U}}_{{0}}}} )^H}{\bf{N}}_t$ and ${{\bf{U}}_0}^H{\bf{Y}}_{S,R}{\rm{ = }}\textbf{0}$ can be discarded. By using the LS method~\cite{Lifu1993}, ${\bf{\Delta }}{{\bf{U}}_0}$ can be presented as
\begin{equation}\label{equ:delta_U0}
	{\bf{\Delta }}{{\bf{U}}_0} =  - {{\bf{U}}_s}{{\bf{\Sigma }}_s^{ - 1}}{\bf{V}}_s^H{[ {{\bf{N}}_t} ]^H}{{\bf{U}}_0}.
\end{equation}
The MUSIC 2D angle estimation result is distorted by the  noise perturbation, which is expressed as	${{\bf{\tilde p}}_k} = {{\bf{p}}_k} + {\bf{\Delta }}{{\bf{p}}_k}$,
where ${{\bf{p}}_k}$ is the actual value of AoA. Apply Taylor series decomposition to $f_a( {{{{\bf{\tilde p}}}_k};{{{\bf{\tilde U}}}_0}} )$ in \eqref{equ:spectrum function}, and take the first three terms. Applying first-order derivative to the truncated Taylor series, we have
\begin{equation}\label{equ:f_p}
	\frac{{\partial f_a( {{{\bf{\tilde p}}_k};{{{\bf{\tilde U}}}_0}} )}}{{\partial {\bf{p}}}} \buildrel\textstyle.\over= \frac{{\partial f_a( {{{\bf{p}}_k};{{{\bf{\tilde U}}}_0}} )}}{{\partial {\bf{p}}}} + \frac{{{\partial ^2}f_a( {{{\bf{p}}_k};{{{\bf{\tilde U}}}_0}} )}}{{{\partial ^2}{\bf{p}}}}\Delta {{\bf{p}}_k}.
\end{equation}

By setting \eqref{equ:f_p} to be 0, we can obtain
\begin{equation}\label{equ:delta_pk}
	\Delta {{\bf{p}}_k} =  - {{\bf{H}}_{\bf{p}}}^{ - 1}( {{{\bf{p}}_k};{{{\bf{\tilde U}}}_0}} ){{\bf{G}}_\textbf{p}}( {{{\bf{p}}_k};{{{\bf{\tilde U}}}_0}} ),
\end{equation}
where ${{\bf{H}}_{\bf{p}}}( {{{\bf{p}}_k};{{{\bf{\tilde U}}}_0}} ) = \frac{{{\partial ^2}f_a( {{{\bf{p}}_k};{{{\bf{\tilde U}}}_0}} )}}{{{\partial ^2}{\bf{p}}}} \in \mathbb{C}^{2 \times 2}$ is the Hessian matrix of $f_a$, and ${{\bf{G}}_\textbf{p}}( {{{\bf{p}}_k};{{{\bf{\tilde U}}}_0}} ) = \frac{{\partial f_a( {{{\bf{p}}_k};{{{\bf{\tilde U}}}_0}} )}}{{\partial {\bf{p}}}} \in \mathbb{C}^{2 \times 1}$ is the gradient vector of $f_a$.

With the perturbation expression, we can obtain
\begin{equation}\label{equ:Gp_expression}
	{{\bf{G}}_\textbf{p}}( {{{\bf{p}}_k};{{{\bf{\tilde U}}}_0}} ) = {{\bf{G}}_\textbf{p}}( {{{\bf{p}}_k};{{\bf{U}}_0}} ) + \Delta {{\bf{G}}_\textbf{p}},
\end{equation}
\begin{equation}\label{equ:Hp_expression}
	{{\bf{H}}_{\bf{p}}}( {{{\bf{p}}_k};{{{\bf{\tilde U}}}_0}} ) = {{\bf{H}}_{\bf{p}}}( {{{\bf{p}}_k};{{\bf{U}}_0}} ) + \Delta {{\bf{H}}_{\bf{p}}}.
\end{equation}

From \eqref{equ:spectrum function}, we have
\begin{equation}\label{equ:G_p_gradient}
	{{\bf{G}}_\textbf{p}}( {{{\bf{p}}_k};{\bf{U}}} ){\rm{ = }}\frac{{\partial f_a( {{\bf{p}};{\bf{U}}} )}}{{\partial {\bf{p}}}} = 2{\mathop{\rm Re}\nolimits} \{ {{\bf{a}}_{\bf{p}}^{( 1 )}{{( {\bf{p}} )}^H}{\bf{U}}{{\bf{U}}^H}{\bf{a}}( {\bf{p}} )} \},
\end{equation}
\begin{equation}\label{equ:Hp}
		\text{vec}[ {{{\bf{H}}_{\bf{p}}}( {{\bf{p}};{\bf{U}}} )} ] 
		= 2{\mathop{\rm Re}\nolimits} \left\{ \begin{array}{l}
			\text{vec}[ {{\bf{a}}_{\bf{p}}^{( 1 )}{{( {\bf{p}} )}^H}{\bf{U}}{{\bf{U}}^H}{\bf{a}}_{\bf{p}}^{( 1 )}( {\bf{p}} )} ]
			+ {\bf{a}}_{\bf{p}}^{( 2 )}{( {\bf{p}} )^H}{\bf{U}}{{\bf{U}}^H}{\bf{a}}( {\bf{p}} )
		\end{array} \right\},
\end{equation}
where $\text{vec}(  \cdot  )$ is to vectorize a matrix, ${\bf{a}}_{\bf{p}}^{( 1 )}( {\bf{p}} )$ and ${\bf{a}}_{\bf{p}}^{( 2 )}( {\bf{p}} )$ are the first-order and second-order derivatives of ${\bf{a}}( {\bf{p}} )$ over ${\bf{p}}$, respectively, which can be derived from~\eqref{equ:steeringVec}.

Since ${{\bf{U}}_0}^H{{\bf{A}}_{S,RX}} = 0$, we can obtain
\begin{equation}\label{equ:G_p_0}
	{{\bf{G}}_{\bf{p}}}( {{{\bf{p}}_k};{{\bf{U}}_0}} ) = {\bf{0}},
\end{equation}
\begin{equation}\label{equ:H_p_0}
	{{\bf{H}}_{\bf{p}}}( {{{\bf{p}}_k};{{\bf{U}}_0}} ) \!=\! 2{\mathop{\rm Re}\nolimits} \{ {{\bf{a}}_{\bf{p}}^{( 1 )}{{( {{{\bf{p}}_k}} )}^H}{{\bf{U}}_0}{{\bf{U}}_0}^H{\bf{a}}_{\bf{p}}^{( 1 )}( {{{\bf{p}}_k}} )} \}.
\end{equation}
We use ${{\bf{H}}_{{\bf{p0}}}}$ to represent ${{\bf{H}}_{\bf{p}}}( {{{\bf{p}}_k};{{\bf{U}}_0}} )$. By substituting \eqref{equ:G_p_0} and \eqref{equ:H_p_0} into \eqref{equ:Gp_expression} and \eqref{equ:Hp_expression}, \eqref{equ:delta_pk} can be rewritten as
\begin{equation}\label{equ:delta_pk_last}
		\Delta {{\bf{p}}_k}	=  - {( {{{\bf{H}}_{{\bf{p0}}}} + \Delta {{\bf{H}}_{\bf{p}}}} )^{ - 1}}( {\Delta {{\bf{G}}_{{\textbf{p}}}}} )
		= - \left( \begin{array}{l}
			{\bf{I}} - {{\bf{H}}_{{\bf{p0}}}}^{ - 1}\Delta {{\bf{H}}_{\bf{p}}} + \\
			{( {{{\bf{H}}_{{\bf{p0}}}}^{ - 1}\Delta {{\bf{H}}_{\bf{p}}}} )^2} + ...
		\end{array} \right){{\bf{H}}_{{\bf{p0}}}}^{ - 1}\Delta {{\bf{G}}_{{\textbf{p}}}}.
\end{equation}
Discarding the perturbation terms that are higher than second-order in \eqref{equ:delta_pk_last}, we can rewrite \eqref{equ:delta_pk_last} as 
\begin{equation}\label{equ:delta_pk_lastlast}
	\Delta {{\bf{p}}_k} =  - {{\bf{H}}_{{\bf{p0}}}}^{ - 1}\Delta {{\bf{G}}_{{\textbf{p}}}},
\end{equation}
where the perturbation expression of $\Delta {{\bf{G}}_\textbf{p}}$ is derived in \textbf{Appendix} \ref{Expressions:G} as
\begin{equation}\label{equ:delta_G_p_last}
	\Delta {{\bf{G}}_{\bf{p}}} \!\!=\!\! 2{\mathop{\rm Re}\nolimits} \{\! { - {\bf{a}}_{\bf{p}}^{( 1 )}{{( {{{\bf{p}}_k}} )}^H}{{\bf{U}}_0}{{\bf{U}}_0}^H{{[ {{\bf{N}}_t} ]}^H}{{\bf{V}}_s}{{\bf{\Sigma }}_s^{ - 1}}{{\bf{U}}_s}^H{\bf{a}}( {{{\bf{p}}_k}} )} \!\},
\end{equation}
By substituting \eqref{equ:delta_G_p_last} and \eqref{equ:H_p_0} into \eqref{equ:delta_pk_lastlast}, we can obtain $\Delta {{\bf{p}}_k}$ as shown in \eqref{equ:delta_pk_lastlastlast}.
\begin{equation}\label{equ:delta_pk_lastlastlast}
	\Delta {{\bf{p}}_k} = {[ {{\mathop{\rm Re}\nolimits} \{ {{\bf{a}}_{\bf{p}}^{( 1 )}{{( {{{\bf{p}}_k}} )}^H}{{\bf{U}}_0}{{\bf{U}}_0}^H{\bf{a}}_{\bf{p}}^{( 1 )}( {{{\bf{p}}_k}} )} \}} ]^{ - 1}}{\mathop{\rm Re}\nolimits} \{ {{\bf{a}}_{\bf{p}}^{( 1 )}{{( {{{\bf{p}}_k}} )}^H}{{\bf{U}}_0}{{\bf{U}}_0}^H[ {{\bf{N}}_t} ]{{\bf{V}}_s}{{\bf{\Sigma }}_s}^{ - 1}{{\bf{U}}_s}^H{\bf{a}}( {{{\bf{p}}_k}} )} \}.
\end{equation}
The MSE of angle estimation can be expressed as
\setcounter{equation}{60}
\begin{equation}\label{equ:MSE_delta_G_p_last}
	MSE( {{{\bf{p}}_k}} ) = E\{ {diag( {\Delta {{\bf{p}}_k}{{[ {\Delta {{\bf{p}}_k}} ]}^H}} )} \}.
\end{equation}

\subsection{Analysis of Range and Doppler Detection MSE}
\subsubsection{Analysis of Range Detection MSE}
The noisy signal for the range estimation, as shown in \eqref{equ:Y_SSU_As}, is rewritten as
\begin{equation}\label{equ:Y_SS_U}
	{\bf{\bar H}}_S = {\bf{\bar H}}_{S,p} + {\bf{W}}_{tr},
\end{equation}
where ${\bf{\bar H}}_{S,p} = {{\bf{A}}_{\bf{r}}}{\bf{S}}_{r,s}$ is the useful signal.
The singular value decomposition of ${\bf{\bar H}}_{S,p}$ is 
\begin{equation}\label{equ:svd_of_Y_SS_p}
	{\bf{\bar H}}_{S,p} \!\!=\!\! \left[ {{{\bf{U}}_{r,s}},{{\bf{U}}_{r,0}}} \right]\left(\!\! {\begin{array}{*{20}{c}}
			{{{\bf{\Sigma }}_{r,s}}}&{\bf{0}}\\
			{\bf{0}}&{\bf{0}}
	\end{array}} \!\right)\left[\! {\begin{array}{*{20}{c}}
			{{\bf{V}}_{r,s}^H}\\
			{{\bf{V}}_{r,0}^H}
	\end{array}} \!\right] \!\! =\!\! {{\bf{U}}_{r,s}}{{\bf{\Sigma }}_{r,s}}{\bf{V}}_{r,s}^H,
\end{equation}
where ${{\bf{U}}_{r,s}}$ and ${{\bf{U}}_{r,0}}$ are orthogonal unitary matrices, and ${{\bf{U}}_{r,0}}^H{\bf{\bar H}}_{S,p} = {\bf{0}}$. We further obtain ${{\bf{U}}_{r,0}}^H{{\bf{A}}_{\bf{r}}} = {\bf{0}}$. 

By treating ${\bf{W}}_{tr}$ as a perturbation term, ${\bf{\bar H}}_S$ can be decomposed as
\begin{equation}\label{equ:svd_of_Y_SS_U}
	{\bf{\bar H}}_S = \left[ {{{{\bf{\tilde U}}}_{r,s}},{{{\bf{\tilde U}}}_{r,0}}} \right]\left( {\begin{array}{*{20}{c}}
			{{{{\bf{\tilde \Sigma }}}_{r,s}}}&{\bf{0}}\\
			{\bf{0}}&{{\bf{\Delta }}{{\bf{\Sigma }}_{r,{\bf{0}}}}}
	\end{array}} \right)\left[ {\begin{array}{*{20}{c}}
			{{\bf{\tilde V}}_{r,s}^H}\\
			{{\bf{\tilde V}}_{r,0}^H}
	\end{array}} \right],
\end{equation}
where ${{\bf{\tilde \Sigma }}_{r,{\bf{0}}}} = {\bf{\Delta }}{{\bf{\Sigma }}_{r,{\bf{0}}}}$, and ${{\bf{\tilde U}}_{r,0}} = {{\bf{U}}_{r,0}} + \Delta {{\bf{U}}_{r,0}}$. Because ${{\bf{\tilde U}}_{r,s}}$ and ${{\bf{\tilde U}}_{r,0}}$ are orthogonal unitary matrices, we have ${{\bf{\tilde U}}_{r,0}}^H{\bf{\bar H}}_S{\rm{ = }}{\bf{\Delta }}{{\bf{\Sigma }}_{r,{\bf{0}}}}{\bf{\tilde V}}_{r,0}^H$. In the high SINR regime, solving the perturbation problem is equivalent to seeking the optimal $\Delta {{\bf{U}}_{r,0}}$ to minimize $\| {{{{\bf{\tilde U}}}_{r,0}}^H{\bf{\bar H}}_S} \|_2$ with the constraint ${{\bf{\tilde U}}_{r,0}}^H{{\bf{\tilde U}}_{r,0}} = {\bf{I}}$. By substituting \eqref{equ:Y_SS_U} and ${{\bf{\tilde U}}_{r,0}} = {{\bf{U}}_{r,0}} + \Delta {{\bf{U}}_{r,0}}$ into the problem, then discarding the term ${{\bf{U}}_{r,0}}^H{\bf{\bar H}}_{S,p} = {\bf{0}}$ and the second-order perturbation $\Delta {{\bf{U}}_{r,0}}^H{\bf{W}}_{tr}$, we can obtain
\begin{equation}\label{equ:Ur_YSSU}
	\| {{{{\bf{\tilde U}}}_{r,0}}^H{\bf{\bar H}}_S} \|_2 \buildrel\textstyle.\over= \Delta {{\bf{U}}_{r,0}}^H{\bf{\bar H}}_{S,p} + {{\bf{U}}_{r,0}}^H{\bf{W}}_{tr}.
\end{equation}
Using the LS method and substituting \eqref{equ:svd_of_Y_SS_p} into \eqref{equ:Ur_YSSU}, we can obtain $\Delta {{\bf{U}}_{r,0}}$ as
\begin{equation}\label{equ:Ur0}
	\Delta {{\bf{U}}_{r,0}} =  - {{\bf{U}}_{r,s}}{{\bf{\Sigma }}_{r,s}}^{ - 1}{\bf{V}}_{r,s}^H{( {{\bf{W}}_{tr}} )^H}{{\bf{U}}_{r,0}}.
\end{equation}

Next, we derive the expression for the perturbation of range estimation, i.e., $\Delta r = r - {r_k}$, where $r_k$ is the actual value of range, and $r$ is the estimation value.

Apply Taylor series decomposition to \eqref{equ:fr} at $r_k$, and keep the first three terms. Applying the first-order derivative to the truncated series with respect to $r$, we obtain the range perturbation as
\begin{equation}\label{equ:taylor_fr}
	\frac{{\partial f_r( {r;{{{\bf{\tilde U}}}_{r,0}}} )}}{{\partial r}} = \frac{{\partial f_r( {{r_k};{{{\bf{\tilde U}}}_{r,0}}} )}}{{\partial r}} + \frac{{{\partial ^2}f_r( {{r_k};{{{\bf{\tilde U}}}_{r,0}}} )}}{{{\partial ^2}r}}\Delta r.
\end{equation}
Because the Newton descent method identifies the optimal point with $\frac{{\partial f_r( {r;{{{\bf{\tilde U}}}_{r,0}}} )}}{{\partial r}}{\rm{ = }}0$, the range perturbation can be expressed as
\begin{equation}\label{equ:range_perturbation}
	\Delta r =  - {[ {H_r( {r_k;{{{\bf{\tilde U}}}_{r,0}}} )} ]^{ - 1}}G_r( {r_k;{{{\bf{\tilde U}}}_{r,0}}} ),
\end{equation}
where 
\begin{equation}\label{equ:Gr_expression}
	G_r( {r;{\bf{U}}} ) = \frac{{\partial f_r( {r_k;{\bf{U}}} )}}{{\partial r}} = 2{\mathop{\rm Re}\nolimits} [ {{\bf{a}}_r^{( 1 )}{{( r )}^H}{\bf{U}}{{{\bf{U}}}^H}{{\bf{a}}_r}( r )} ],
\end{equation}
and
\begin{equation}\label{equ:Hr_expression}
		H_r( {r;{\bf{U}}} ) = \frac{{{\partial ^2}f_r( {r;{\bf{U}}} )}}{{{\partial ^2}r}}\\
		= 2{\mathop{\rm Re}\nolimits} \left[ \begin{array}{l}
			{\bf{a}}_r^{\left( 2 \right)}{\left( r \right)^H}{\bf{U}}{ {\bf{U}}^H}{{\bf{a}}_r}\left( r \right)
			+ {\bf{a}}_r^{\left( 1 \right)}{\left( r \right)^H}{\bf{U}}{{\bf{U}}^H}{\bf{a}}_r^{\left( 1 \right)}\left( r \right)
		\end{array} \right].
\end{equation}
Using the perturbation form to express $G_r( {r_k;{{{\bf{\tilde U}}}_{r,0}}} )$ and ${H_r( {r_k;{{{\bf{\tilde U}}}_{r,0}}} )}$, we have 
\begin{equation}\label{equ:Gr_U0bar}
	G_r( {r_k;{{{\bf{\tilde U}}}_{r,0}}} ) = G_r( {r_k;{{\bf{U}}_{r,0}}} ) + \Delta G_r,
\end{equation}
and
\begin{equation}\label{equ:Hr_U0bar}
	H_r( {r_k;{{{\bf{\tilde U}}}_{r,0}}} ) = H_r( {r_k;{{\bf{U}}_{r,0}}} ) + \Delta H_r.
\end{equation}
Because ${{\bf{U}}_{r,0}}^H{{\bf{A}}_{\bf{r}}} = 0$, we have
\begin{equation}\label{equ:Gr_rk}
	G_r( {r_k;{{\bf{U}}_{r,0}}} ) = 0,
\end{equation}
and 
\begin{equation}\label{equ:Hr_rk}
	H_r( {{r_k};{{\bf{U}}_{r,0}}} ) \!\!=\!\! 2{\mathop{\rm Re}\nolimits} [ {{\bf{a}}_r^{( 1 )}{{( {{r_k}} )}^H}{{\bf{U}}_{r,0}}{{( {{{\bf{U}}_{r,0}}} )}^H}{\bf{a}}_r^{( 1 )}( {{r_k}} )} ]\!\! =\!\! H_{r0},
\end{equation}
By substituting \eqref{equ:Gr_rk} and \eqref{equ:Hr_rk} into \eqref{equ:Gr_U0bar} and \eqref{equ:Hr_U0bar}, respectively, \eqref{equ:range_perturbation} becomes
\begin{equation}\label{equ:delta_r}
		\Delta r_k	=  - {\{ {H_{r0}[ {1 + {{( {H_{r0}} )}^{ - 1}}\Delta H_r} ]} \}^{ - 1}}\Delta G_r\\
		\buildrel\textstyle.\over=  - {( {H_{r0}} )^{ - 1}}\Delta G_r,
\end{equation}
where the last equation is obtained by discarding the second-order perturbation terms.

The perturbation expression of $\Delta G_r$ is derived in \textbf{Appendix} \ref{Expressions:G}, given by
\begin{equation}\label{equ:delta_Gr}
		\Delta G_r = 2{\mathop{\rm Re}\nolimits} [ {{\bf{a}}_r^{( 1 )}{{( r )}^H}( {{{\bf{U}}_{r,0}}\Delta {{\bf{U}}_{r,0}}^H} ){{\bf{a}}_r}( r )} ],
\end{equation}
By substituting \eqref{equ:Hr_rk}, \eqref{equ:delta_Gr}, and \eqref{equ:Ur0} into \eqref{equ:delta_r}, we obtain
\begin{equation}\label{equ:delta_r_last}
	\Delta r_k \!\!=\!\! \frac{{{\mathop{\rm Re}\nolimits} [ {{\bf{a}}_r^{\!( 1 )}{{\!( {{r_k}} )}^H}\!{{\bf{U}}_{r,0}}{{\bf{U}}_{r,0}}^H \!{\bf{W}}_{tr}{{\bf{V}}_{r,s}}{{\bf{\Sigma }}_{r,s}}^{\!\! - 1}\!{{\bf{U}}_{r,s}}^H{{\bf{a}}_r}( {{r_k}} )} ]}}{{{\mathop{\rm Re}\nolimits} [ {{\bf{a}}_r^{( 1 )}{{( {{r_k}} )}^H}{{\bf{U}}_{r,0}}{{( {{{\bf{U}}_{r,0}}} )}^H}{\bf{a}}_r^{( 1 )}( {{r_k}} )} ]}},
\end{equation}
where ${{\bf{a}}_r}( {{r_k}} )$ is given in \eqref{equ:range_steering}, and ${\bf{a}}_r^{( 1 )}{( {{r_k}} )}$ is given in \eqref{equ:first_derivative_a}.

The MSE of the MUSIC-based JCS range estimation can be expressed as
\begin{equation}\label{equ:MSE_r}
	MSE( r ) = E[ {\Delta {r_k^2}} ].
\end{equation}
\subsubsection{Analysis of Doppler Detection MSE}
Similar to the range estimation, the perturbation of Doppler estimation can be derived as
\begin{equation}\label{equ:delta_f}
	\Delta {f_d}\!\! =\!\! \frac{{{\mathop{\rm Re}\nolimits} [ {{\bf{a}}_f^{( 1 )}{{\! ( {{f_k}} )}^H}\!{{\bf{U}}_{f,0}}\!{{\bf{U}}_{f,0}}^H \!{\bf{W}}_{tf}{{\bf{V}}_{f,s}}{{\bf{\Sigma }}_{f,s}}^{\!\! - 1}\!{{\bf{U}}_{f,s}}^{\!\! H} \!{{\bf{a}}_f} \!( {{f_k}} )} ]}}{{{\mathop{\rm Re}\nolimits} [ {{\bf{a}}_f^{( 1 )}{{( {{f_k}} )}^H}{{\bf{U}}_{f,0}}{{( {{{\bf{U}}_{f,0}}} )}^H}{\bf{a}}_f^{( 1 )}( {{f_k}} )} ]}},
\end{equation}
where $f_k$ is the real Doppler value, ${{{\bf{a}}_f}( f_k )}$ is given in \eqref{equ:doppler_steering}, and ${\bf{a}}_f^{( 1 )}( f_k )$ is given in \eqref{equ:first_derivative_a}. Furthermore, the perturbation of the radial velocity estimation is
\begin{equation}\label{equ:delta_v}
	\Delta v = \lambda \Delta f_d.
\end{equation}

\subsection{Analysis of Location MSE}
The location of the target can be obtained after the AoA, ${{\bf{p}}_k} = \left( {{\varphi _k},{\theta _k}} \right)$, and the range, ${r_k}$, are detected. The expression for the actual location is given by
\begin{equation}\label{equ:ploc}
		{{\bf{p}}_{loc}}\!( {{r_k},{\varphi _k},{\theta _k}} )\! = \! ( {{x_k},{y_k},{z_k}} ) \\
		\!= \! ( {{r_k}\sin {\theta _k}\cos {\varphi _k},r_k\sin {\theta _k}\sin {\varphi _k},r_k\cos {\theta _k}} ).
\end{equation}
With the AoA and range estimation perturbation, $\Delta {{\bf{p}}_k} = ( {\Delta {\varphi _k},\Delta {\theta _k}} )$ and $\Delta {r_k}$, the location of the target is 
\begin{equation}\label{equ:ploc_perturbation}
			{{\bf{p}}_{loc}}( {{r_k} + \Delta {r_k},{\varphi _k} + \Delta {\varphi _k},{\theta _k} + \Delta {\theta _k}} )\\
			= \left[ \begin{array}{l}
				( {{r_k} + \Delta {r_k}} )\sin ( {{\theta _k} + \Delta {\theta _k}} )\cos ( {{\varphi _k} + \Delta {\varphi _k}} ),\\
				( {{r_k} + \Delta {r_k}} )\sin ( {{\theta _k} + \Delta {\theta _k}} )\sin ( {{\varphi _k} + \Delta {\varphi _k}} ),\\
				( {{r_k} + \Delta {r_k}} )\cos ( {{\theta _k} + \Delta {\theta _k}} )
			\end{array} \right],
\end{equation}
Comparing \eqref{equ:ploc} with \eqref{equ:ploc_perturbation} and discarding the second-order perturbation, we can represent the perturbation of $x$, $y$, and $z$ axes coordinates as
\begin{equation}\label{equ:delta_x}
	\Delta x \buildrel\textstyle.\over= \Delta {r_k}\sin {\theta _k}\cos {\varphi _k} + {r_k}\left(\! \begin{array}{l}
		\Delta {\theta _k}\cos {\theta _k}\cos {\varphi _k}
		- \Delta {\varphi _k}\sin {\theta _k}\sin {\varphi _k}
	\end{array} \!\right),
\end{equation}
\begin{equation}\label{equ:delta_y}
	\Delta y \buildrel\textstyle.\over= \Delta {r_k}\sin {\theta _k}\sin {\varphi _k} + {r_k}\left(\! \begin{array}{l}
		\Delta {\varphi _k}\sin {\theta _k}\cos {\varphi _k}
		+ \Delta {\theta _k}\cos {\theta _k}\sin {\varphi _k}
	\end{array} \!\right),
\end{equation}
and
\begin{equation}\label{equ:delta_z}
	\Delta z \buildrel\textstyle.\over= \Delta {r_k}\cos {\theta _k} - {r_k}\Delta {\theta _k}\sin {\theta _k}.
\end{equation}
Finally, the location error can be expressed as
\begin{equation}\label{equ:ploc_delta}
	E\{ {\| {\Delta {{\bf{p}}_{loc}}} \|_2^2} \} = E\{ {{{( {\Delta x} )}^2} + {{( {\Delta y} )}^2} + {{( {\Delta z} )}^2}} \}.
\end{equation}

{\color{blue}
\subsection{Cramer–Rao bound of JCS Sensing}
We further derive the Cramer–Rao bound (CRB) to characterize the minimum lower bound for sensing. Based on the signal model presented in Section~\ref{sec:DL Echo Sensing Received Signal}, the echo signal of the $l$th target received by the $(p,q)$th antenna element at the $n$th subcarrier of the $m$th OFDM symbol is
\begin{equation}\label{equ:CRB_RX_signal}
	y_{S,n,m,l}^{p,q} \!\!=\!\! \sqrt {P_t} d_{n,m}{\alpha _{S,n,m,l}}\chi _{TX,l}{a_{p,q}}( {{{\bf{p}}_l}} ) + n_{S,n,m}^{p,q} + x_{S,n,m}^{p,q},
\end{equation} 
where ${\alpha _{S,n,m,l}} = {b_{S,l}}{e^{j2\pi mT_s2{v_{s,l}}/\lambda }}{e^{ - j2\pi n\Delta {f}2r{_{s,l}}/c}}$ is given as~\eqref{equ:alpha_S}, $v_{s,l} = v_{r,l,1}$ and $r_{s,l} = d_{l,1}$ are the radial relative velocity and distance between BS and the $l$th target, respectively; ${a_{p,q}}\left( {{{\bf{p}}_l}} \right)$ is given in~\eqref{equ:phase_difference}, ${{\bf{p}}_l} = {\bf{p}}_{TX,l} = ({\varphi _l},{\theta _l})$ is the 2D AoA of the $l$th target; $\chi _{TX,l}$ is the transmitting BF gain; $n_{S,n,m}^{p,q}$ and $x_{S,n,m}^{p,q}$ are the noise and interference at the $(p,q)$th antenna element. Let $n_{S,n,m}^{X,p,q} \triangleq n_{S,n,m}^{p,q} + x_{S,n,m}^{p,q}$, then $n_{S,n,m}^{X,p,q}$ is independent and identically distributed, following $\mathcal{CN}(0, \sigma _W^2)$, where $\sigma _W^2 = {P_{IS}} + \sigma _N^2$. 

Let ${\bf{\psi }} = \left( {{r_{s,l}},{v_{s,l}},{\varphi _l},{\theta _l}} \right)$ be the set of estimation parameters. Then, the distribution of $y_{S,n,m,l}^{p,q}$ is 
\begin{equation}\label{equ:possibility_psi}
	p( {y;{\bf{\psi }}} ) \!\! = \!\! \frac{1}{{\pi \sigma _W^2}}{e^{ - \| {y - \sqrt {P_t} d_{n,m}{\alpha _{S,n,m,l}}\chi _{TX,l}{a_{p,q}}( {{{\bf{p}}_l}} )} \|_2^2/\sigma _W^2}},
\end{equation}
Because there are $N_cM_s{P_t}{Q_t}$ independent symbols used for estimation, the joint distribution of these symbols is 
\begin{equation}\label{equ:possibility_y_vec}
	p\left( {{\bf{y}};{\bf{\psi }}} \right) = \rho {e^{ - \sum\limits_{(n,m,p,q)}^{N_cM_s{P_t}{Q_t}} {\| {y_{n,m}^{p,q} - {s_{n,m}}{\alpha _{S,n,m,l}}{a_{p,q}}( {{{\bf{p}}_l}} )} \|_2^2/\sigma _W^2} }},
\end{equation}
where $\rho  = {( {\frac{1}{{\pi \sigma _W^2}}} )^{N_cM_s{P_t}{Q_t}}}$, and ${s_{n,m}} = \sqrt {P_t} d_{n,m}\chi _{TX,l}$. Note that $d_{n,m}$ is independent and identically distributed with $E( {\| {d_{n,m}} \|_2^2} ) = 1$. According to \cite{Levy2008Principles, CRB2010}, the CRB of ${\psi _i}$, ${\psi _i} \in \left( {{r_{s,l}},{v_{s,l}},{\varphi _l},{\theta _l}} \right)$, is given by
\begin{equation}\label{equ:CRB_def}
	{C_{{\psi _i}}} =  - {\left\{ {E\left[ {\frac{{{\partial ^2}\ln p\left( {{\bf{y}};{\bf{\psi }}} \right)}}{{{\partial ^2}{\psi _i}}}} \right]} \right\}^{ - 1}}.
\end{equation}
With \eqref{equ:possibility_y_vec} and \eqref{equ:CRB_def}, the sensing CRBs are derived as
\begin{equation} \label{equ:CRB}
	\begin{array}{c}
		{C_{{r_{s,l}}}} = \frac{{{c^2}}}{{32{\pi ^2}{\gamma _S}M_s{P_t}{Q_t}\sum\limits_{n = 0}^{N_c - 1} {{n^2}{{( {\Delta {f}} )}^2}} }},
		{C_{{v_{s,l}}}} = \frac{{{\lambda ^2}}}{{32{\pi ^2}{\gamma _S}N_c{P_t}{Q_t}\sum\limits_{m = 0}^{M_s - 1} {{m^2}{{( {{T}} )}^2}} }},\\
		{C_{{\varphi _l}}} = \frac{{{\lambda ^2}}}{{8{\pi ^2}d_a^2{\gamma _S}N_cM_s\sum\limits_{p,q}^{} {{{\left( {q\cos {\varphi _l}\sin {\theta _l} - p\sin {\varphi _l}\sin {\theta _l}} \right)}^2}} }},
		{C_{{\theta _l}}} = \frac{{{\lambda ^2}}}{{8{\pi ^2}d_a^2{\gamma _S}N_cM_s\sum\limits_{p,q}^{} {{{\left( {p\cos {\varphi _l}\cos {\theta _l} + q\sin {\varphi _l}\cos {\theta _l}} \right)}^2}} }},
	\end{array}
\end{equation}
where $\gamma_S$ is the S-SINR as given in \eqref{equ:gamma_s}. 

\begin{figure*}[!t]
	\centering
	\subfigure[Range detection spectrum.]{\includegraphics[width=0.32\textheight]
		{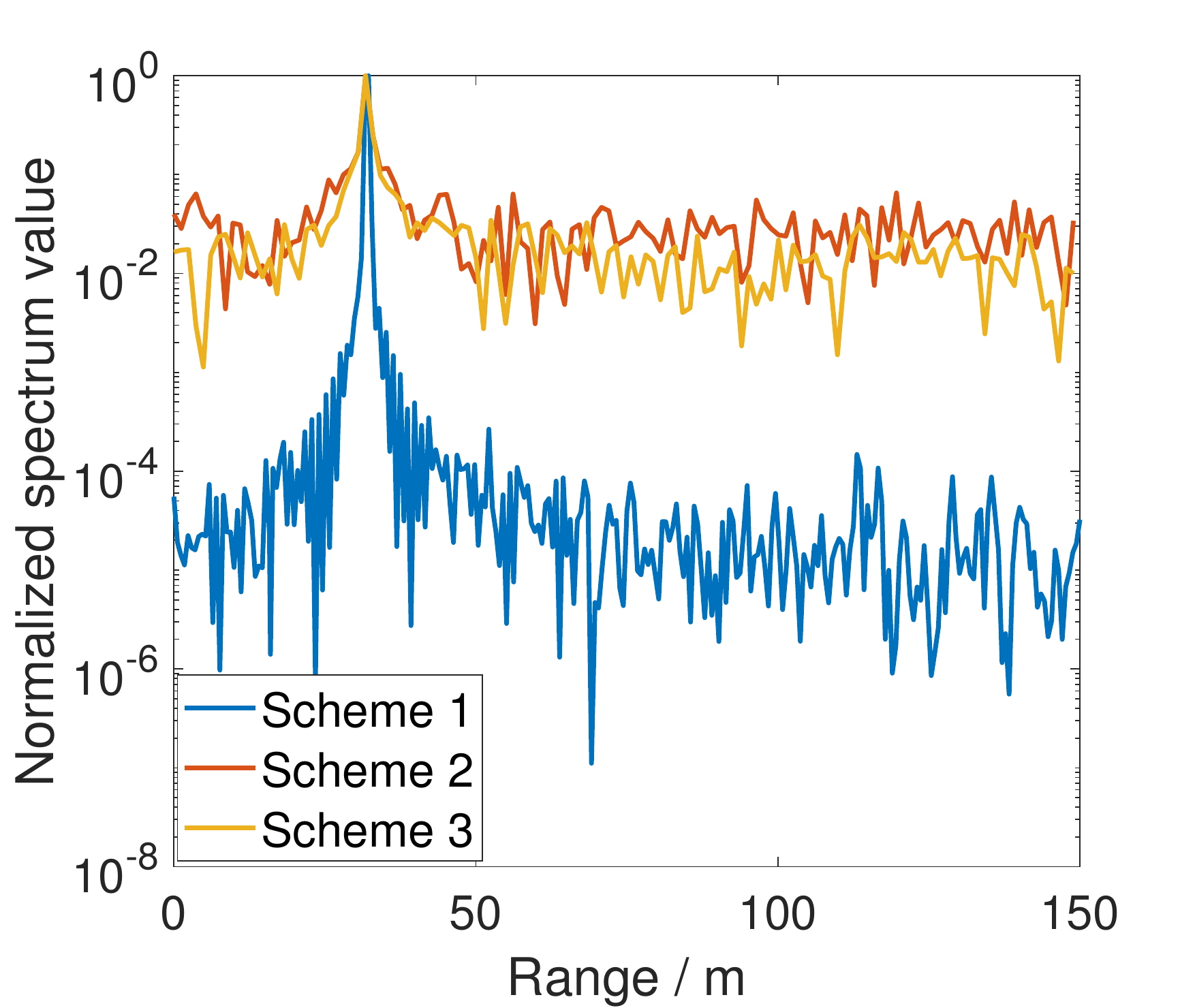}
		\label{figs:Spectrum_range}
	}
	\subfigure[Velocity detection spectrum.]{\includegraphics[width=0.32\textheight]
		{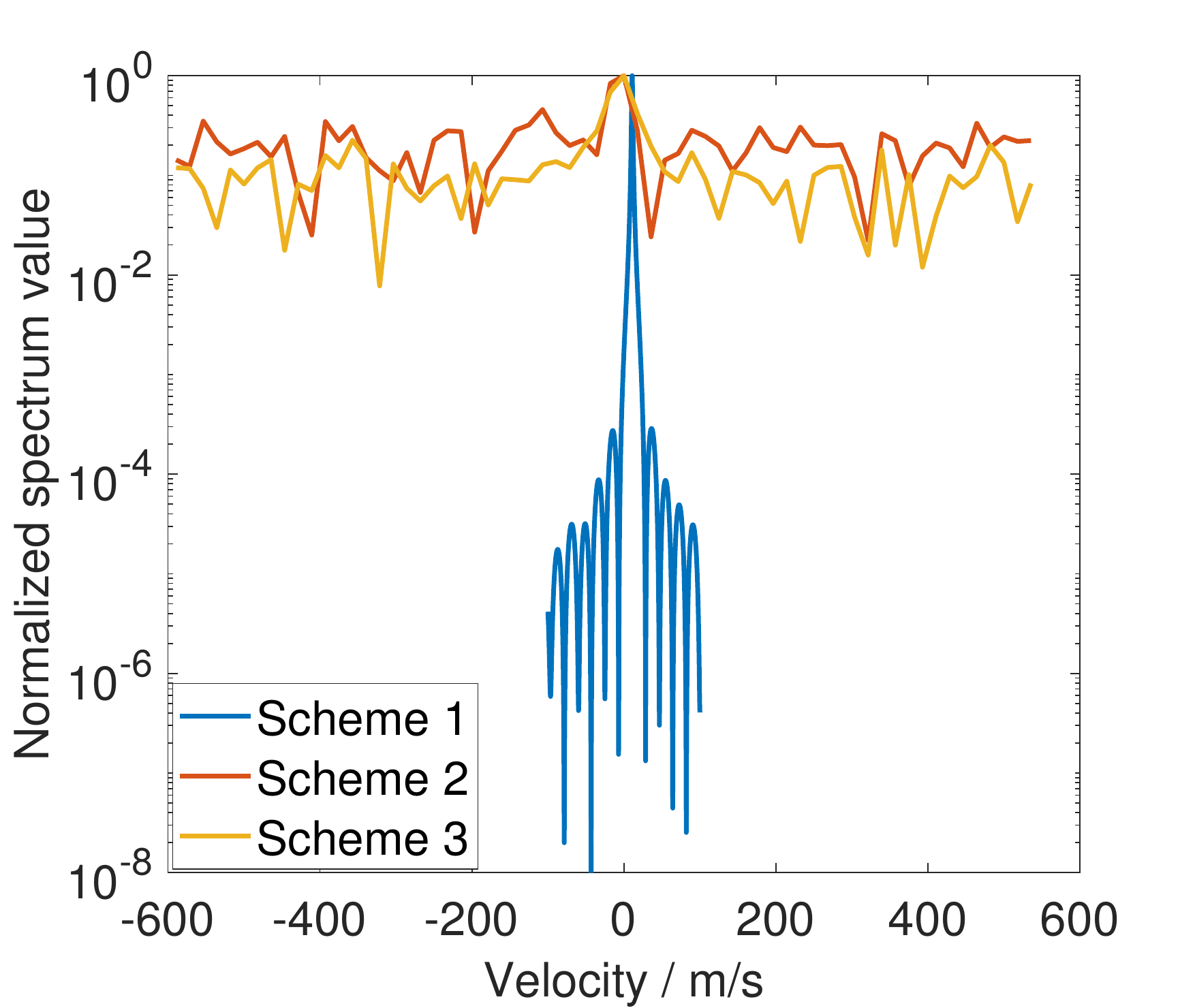}
		\label{figs:Spectrum_velocity}
	}
	\caption{Detection spectra of \textit{schemes} 1, 2 and 3.}
	\label{fig:Detection_Spectrum}
\end{figure*}
}

{\color{blue}
	\subsection{Complexity Analysis and Comparison}\label{sec:cases_complexity}
	In this section, we analyze and compare the complexity of the proposed MUSIC-based JCS method with the conventional FFT-based methods. We consider three schemes: \textit{Scheme} 1 is the proposed MUSIC-based method; \textit{Scheme} 2 is the original FFT-based method in~\cite{Sturm2011Waveform}; and \textit{Scheme} 3 is the Code-division OFDM (CD-OFDM) FFT-based method in~\cite{Chen2021CDOFDM}. 
	
	\textit{Scheme} 1: The main complexity is associated with the eigenvalue decomposition of ${{\bf{R}}_{X,r}}$ and ${{\bf{R}}_{X,f}}$ and the derivation of detection spectra. Therefore, for range and Doppler estimation, the computation complexities are ${\cal{O}}[ {{{( {N_c} )}^3}} ]$ and ${\cal{O}}[ {{{( {M_s} )}^3}} ]$, respectively. Because the MUSIC-based JCS method can work in parallel, the total complexity is ${\cal{O}}\left( {\max \left\{ {{M_s}^3,{N_c}^3} \right\}} \right)$.
	
	\textit{Scheme} 2: The complexity is mainly from two serial FFT operations for the $N_c \times M_s$ echo sensing channel matrix. Therefore, the complexity of \textit{Scheme} 2 is ${\cal{O}}\left( {M_sN_c\log \left( {M_sN_c} \right)} \right)$.
	
	\textit{Scheme} 3: The complexity is mainly from code-division multiplex demodulation and two serial FFT operations for the $N_c \times M_s$ echo sensing channel matrix. Therefore, the complexity of \textit{Scheme} 3 is ${\cal{O}}[ {{{( {N_c} )}^2}M_s + M_sN_c\log ( {M_sN_c} )} ]$.
	
	It can be seen that \textit{Scheme} 2 has the lowest complexity. The complexity of \textit{Scheme} 3 increases due to the additional code-division multiplex processing. The complexity of our proposed MUSIC-based JCS method has the highest complexity to achieve super-resolution detection.
}

\section{Numerical and Simulation Results}\label{sec:JCS result}
In this section, we present extensive simulation results for the proposed MUSIC-based JCS processing method, with comparison to the \textit{Schemes} 2 and 3 as described in~Section~\ref{sec:cases_complexity}, and verify them against the analytical performance bounds derived in Section \ref{sec:JCS performance}. We also compare the BER results of communication demodulation for the proposed JCS CSI enhancement method with those in conventional communication systems. 

\subsection{System Setup}
The system setup largely follows the specification in the 3GPP Vehicles-to-Everything (V2X) applications~\cite{3GPPV2X}. The carrier frequency is 63 GHz, the antenna interval, $d_a$, is half of the wavelength, the sizes of antenna arrays of BS and MUE are $P_t \times Q_t = 8 \times 8$ and $P_r \times Q_r = 1\times 1$, respectively. {\color{blue} The subcarrier interval is $\Delta {f} =$ 480 kHz, the subcarrier number is set to $N_c =$  256, and the number of consecutive OFDM symbols is $M_s = $ 64. Therefore, the bandwidth for JCAS is ${{B  =  }}{N_c}\Delta f = $ 122.88 MHz.} The range and radial velocity resolutions are $\Delta r = \frac{c}{{2B}} = 1.22$ m and $\Delta v = \frac{{\lambda \Delta {f}}}{{2M_s}} = 17.8571$ m/s, respectively~\cite{Sturm2011Waveform}. The variance of the Gaussian noise is $\sigma_N^2 = kFTB = 4.9177\times10^{-12} $ W, where $k = 1.38 \times 10^{-23}$ J/K is the Boltzmann constant, $F = $ 10 is the noise factor, and $T = 290$ K is the standard temperature. The INRs for communication and sensing signals are $\gamma _C^{IN} = \gamma _S^{IN} = $ 3 dB.

{\color{blue}
\begin{figure*}[!t]
	\centering
	\subfigure[AoA detection MSE.]{\includegraphics[width=0.31\textheight]{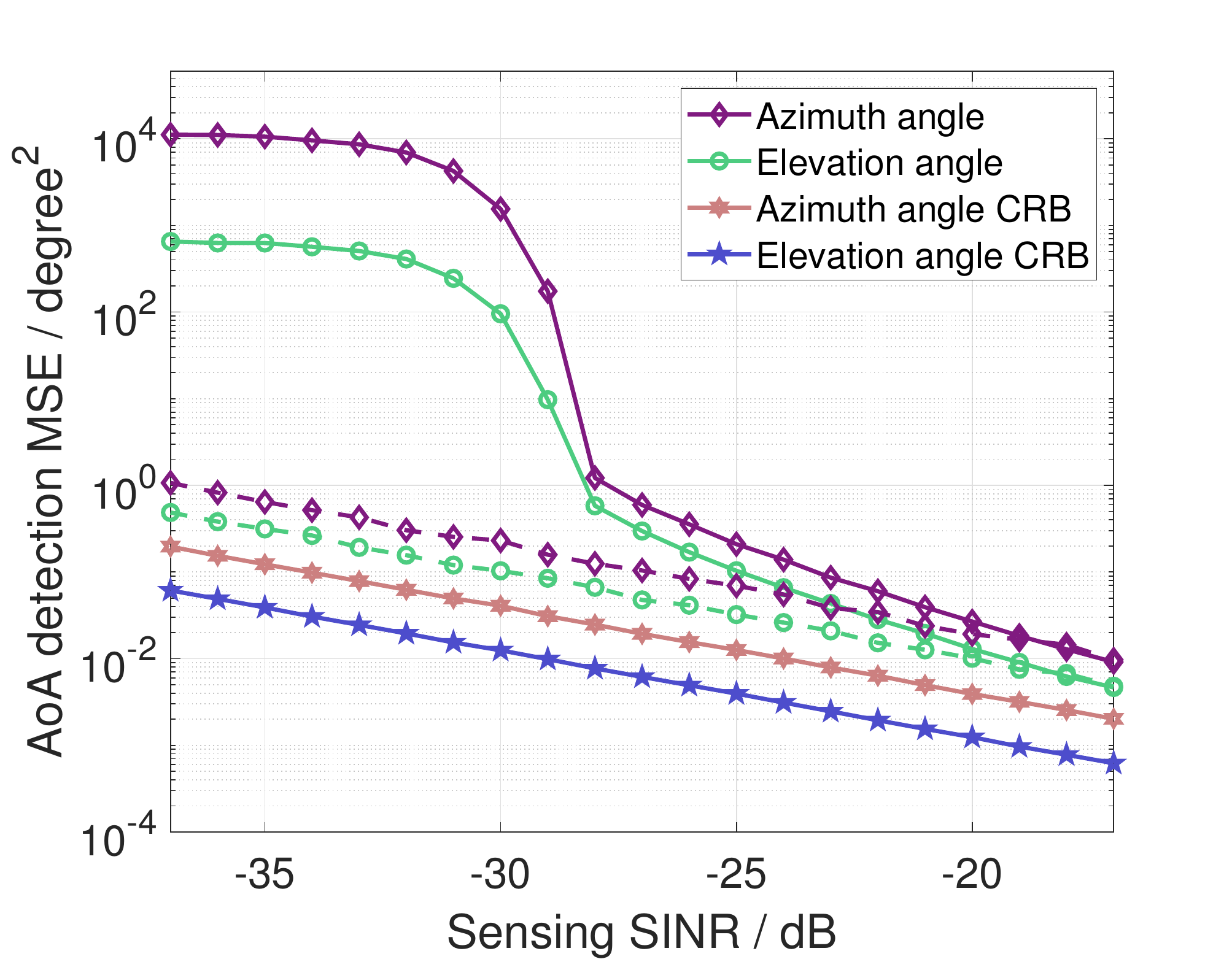}
		\label{figs:phe_theta_Ms64_Nc128}
	}
	\subfigure[Range detection MSE. ]{\includegraphics[width=0.31\textheight]
		{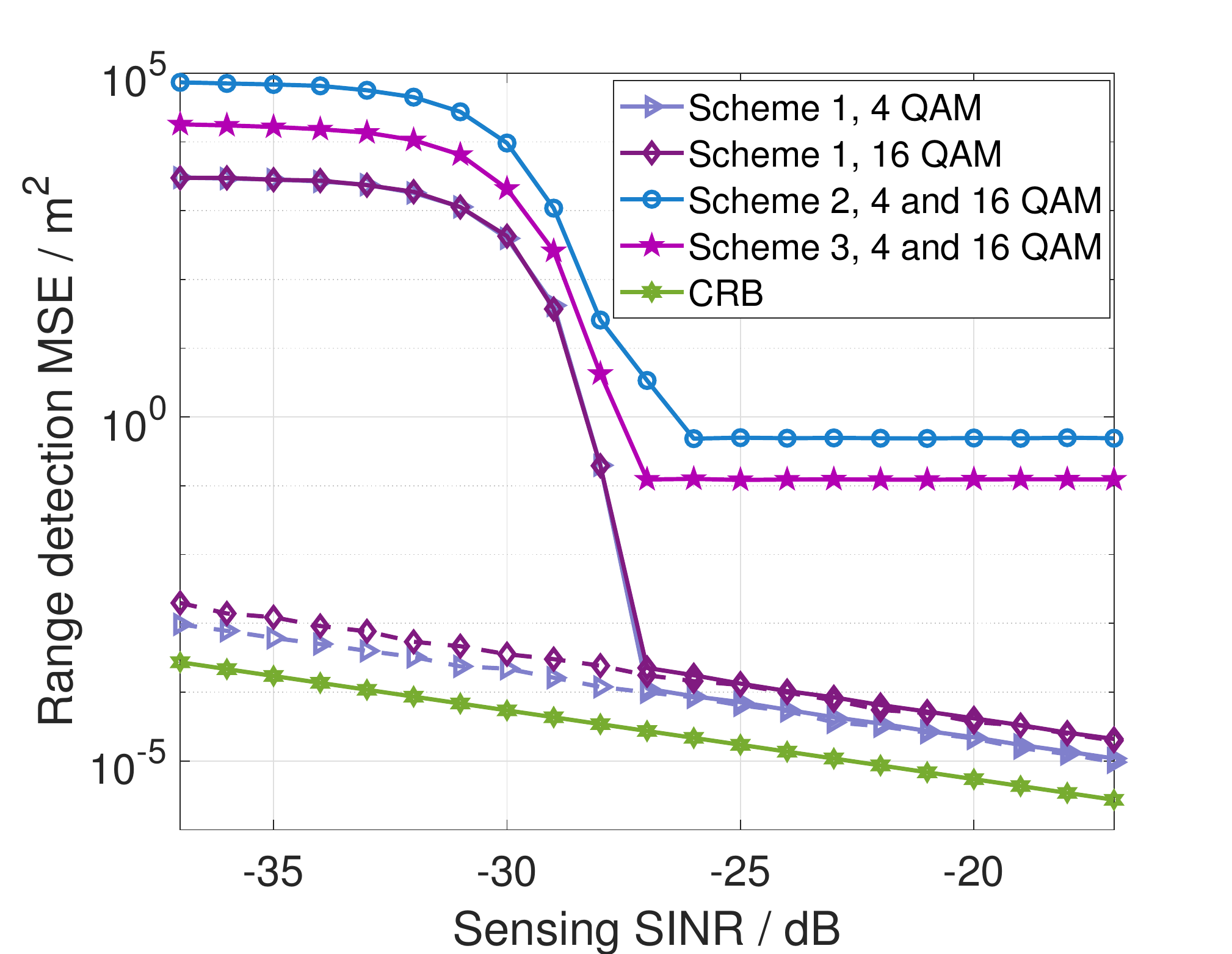}
		\label{figs:Range_Ms64_Nc128}
	}
	\\
	\subfigure[Velocity detection MSE. ]{\includegraphics[width=0.31\textheight]
		{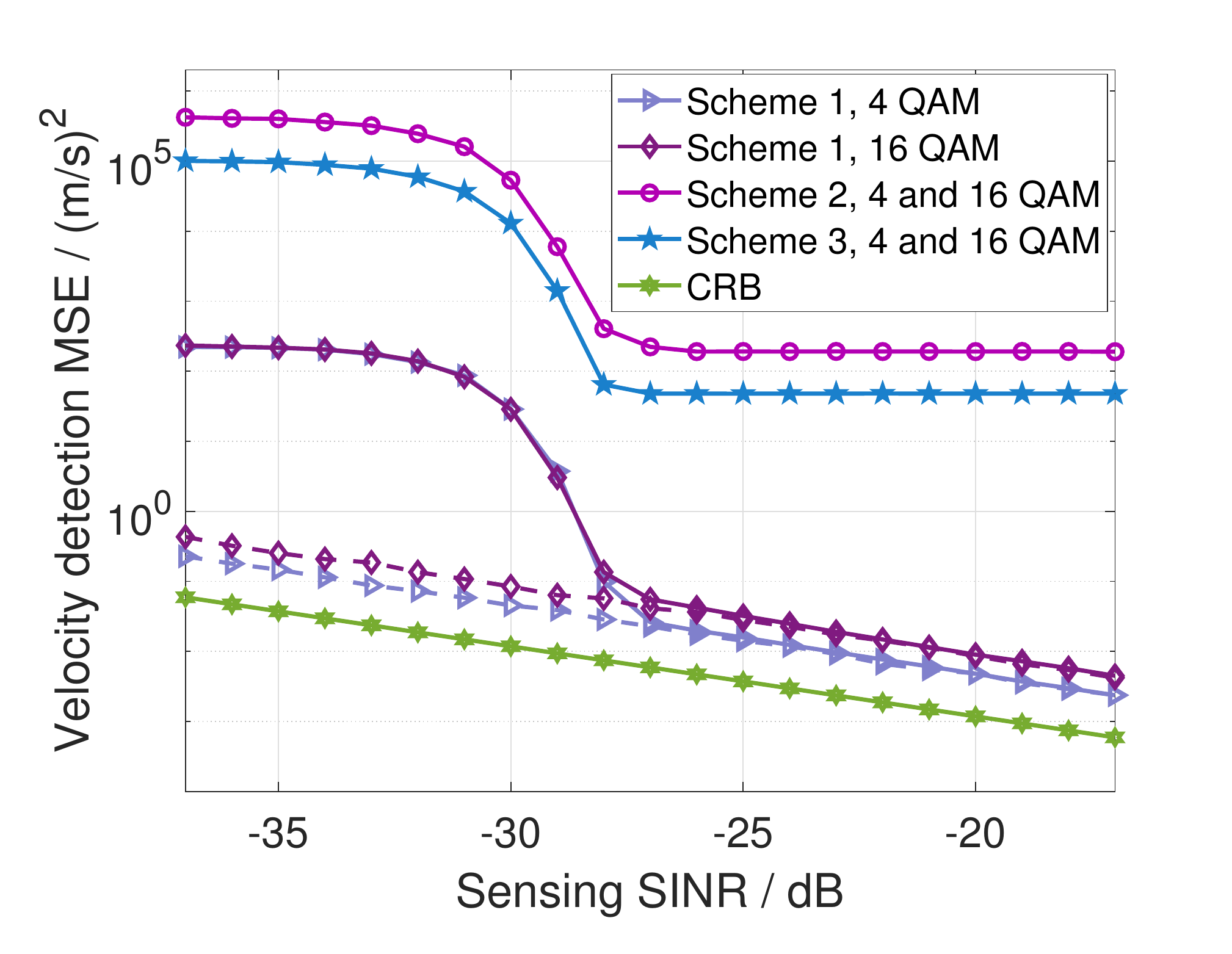}
		\label{figs:Velocity_Ms64_Nc128}
	}
	\subfigure[Location detection MSE.]{\includegraphics[width=0.31\textheight]
		{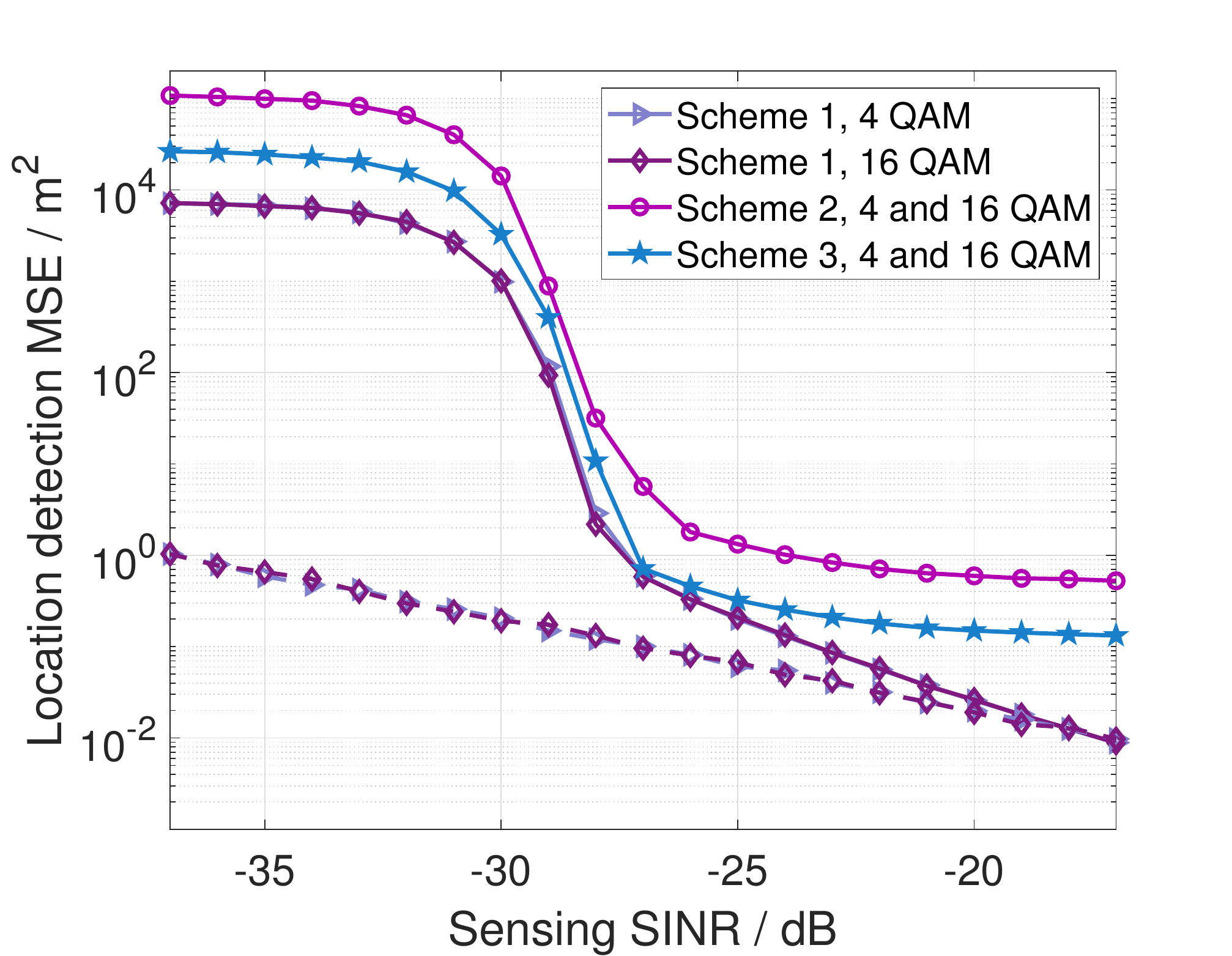}
		\label{figs:location_Ms64_Nc128}
	}\\
	\caption{MSEs for sensing parameter estimation. In Fig.~\ref{figs:phe_theta_Ms64_Nc128}, the solid curves are for the MSEs and CRBs of azimuth and elevation angles obtained in simulation, and the dashed curves are for the numerical ones via the theoretical perturbation results; in Figs.~\ref{figs:Range_Ms64_Nc128}, \ref{figs:Velocity_Ms64_Nc128} and \ref{figs:location_Ms64_Nc128}, the solid curves are for the MSEs and CRBs computed via simulation results, and the dashed curves are for theoretical perturbation results.}
	\label{fig:Detection_Results_Ms64_Nc128}
\end{figure*}
}
Moreover, the location of the BS transmitting array is ${{\bf{p}}_{loc,u}} = (\rm{50, 4.75, 7} )$ m. {\color{blue}  MUE moves on the $x$-axis and its antenna's location is ${{\bf{p}}_{loc,u}} = ({x, 0, 2} )$ m, where $x$ follows uniform distribution from 50 m to 155 m. The scatterer is generated uniformly in a sphere centered at BS with a radius of 100 m. BS is static, while the velocity of MUE is $(-11.11, 0, 0)$ m/s. The reflection factors of the targets are $\sigma _{C\beta ,l}^2 = \sigma _{S\beta ,l}^2 = $ 1. The BS array spins 45 degrees along the $z$-axis and has a downtilt angle of 20 degrees. For each test, the AoAs, ranges, and radial velocities between BS and MUE are then generated from the above parameters, and the JCS communication and echo sensing channel are further generated following the expressions in Section~\ref{subsec:uplink_signal model}. The transmit power of BS for each test, $P_t$, is determined using \eqref{equ:gamma_s} for the given values of S-SINR and INR. 

The MSEs of AoA, range, velocity, and location estimation are defined as the mean values of the squared errors of all the estimates.
}

\subsection{Sensing Performance}
We first demonstrate the sensing spectra of \textit{schemes} 1, 2, and 3. The normalized range spectrum and radial velocity spectrum are shown in Figs. \ref{figs:Spectrum_range} and \ref{figs:Spectrum_velocity}, respectively. The S-SINR is ${\gamma _{S,n,m}} =  - 20$ dB. For range estimation as shown in Fig.~\ref{figs:Spectrum_range}, the peak to sidelobe ratio (PSLR) of \textit{scheme} 1 is about 26 dB. By contrast, the PSLRs of \textit{schemes} 2 and 3 are both around 10 dB. For radial velocity estimation as shown in Fig.~\ref{figs:Spectrum_velocity}, the PSLR of \textit{scheme} 1 is about 33 dB, while the PSLRs of \textit{schemes} 2 and 3 are around 10 dB. The improvement of PSLR of the proposed MUSIC-based JCAS method is credited to the eigenvalue (or singular value) decomposition process, which separates the interference-plus-noise (IN) and signal subspace and reduces the influence of the noise on signal detection.

{\color{blue} 

Fig.~\ref{figs:phe_theta_Ms64_Nc128} presents the AoA estimation MSE of various S-SINRs. With the increase of S-SINR, the AoA estimation MSE decreases as the receiving signal power increases. As the S-SINR is larger than $-$27 dB, the AoA estimation MSE is less than 0.5 square degrees. Since the range of azimuth angle, ${\varphi _k}$, is larger than the elevation angle, ${\theta _k}$, the MSE of ${\varphi_k}$ is larger than ${\theta_k}$ at first. With S-SINR becoming large enough, the MSE of ${\varphi _k}$ approaches that of ${\theta _k}$.

Fig.~\ref{figs:Range_Ms64_Nc128} and Fig.~\ref{figs:Velocity_Ms64_Nc128} demonstrate the range and radial velocity estimation MSEs for \textit{schemes} 1, 2, and 3 under various S-SINRs, respectively. The range and velocity estimation MSEs of \textit{scheme} 3 outperform \textit{scheme} 2 because the code-division multiplex processing in \textit{scheme} 3 can suppress the interference to a certain extent. \textit{scheme} 1 achieves much lower MSEs than both \textit{schemes} 2 and 3, closer to the CRBs in the high SINR regime. This is because the resolutions of \textit{schemes} 2 and 3 are constrained by their FFT-based sensing, with  $({\Delta r})^2 = 1.5$ $m^2$ and ${\left( {\Delta v} \right)^2} = 318 \, {\left( {m/s} \right)^2}$ in this simulation setting. In contrast, our proposed MUSIC-based method can sample the consecutive range and velocity spectra and achieves range and velocity MSEs lower than $10^{-3}$ $m^2$ and $10^{-3}$ $(m/s)^2$, respectively. 
The MSEs for \textit{scheme} 1 is about 25 dB lower than those for \textit{scheme} 3, closer to the range and velocity CRBs. These results demonstrate that the proposed MUSIC-based JCS method achieves super-resolution sensing. Moreover, the theoretical MSEs are shown to be close to the simulation MSEs in the high SINR regime. The higher QAM order results in larger MSEs for \textit{scheme} 1, because the increase of QAM order results in larger transformed noise as can be seen from \eqref{equ:y_n_m_k} and \eqref{equ:Y_SSU}.

Fig.~\ref{figs:location_Ms64_Nc128} shows the location MSE versus S-SINR. With the estimated AoA and range, the location can be determined by \eqref{equ:ploc}. Given the sensing SINR, the MUSIC-based JCS method achieves better location MSE than \textit{scheme} 3. The gaps between \textit{scheme} 1 and \textit{scheme} 3 are not so large in the high SINR regime. This is because the AoA estimation error dominates the location MSE. More specifically, $E\{{({\hat r_{s,l}} - { r_{s,l}})}^2\}$ is smaller than $10^{-1}$ $m^2$, while the error of location as shown in \eqref{equ:delta_x}, \eqref{equ:delta_y}, and \eqref{equ:delta_z} can be much larger than $E\{{({\hat r_{s,l}} - { r_{s,l}})}^2\}$, because they are related to $r_{s,l}$.
}

\begin{figure}[!t]
	\centering
	\includegraphics[width=0.33\textheight]{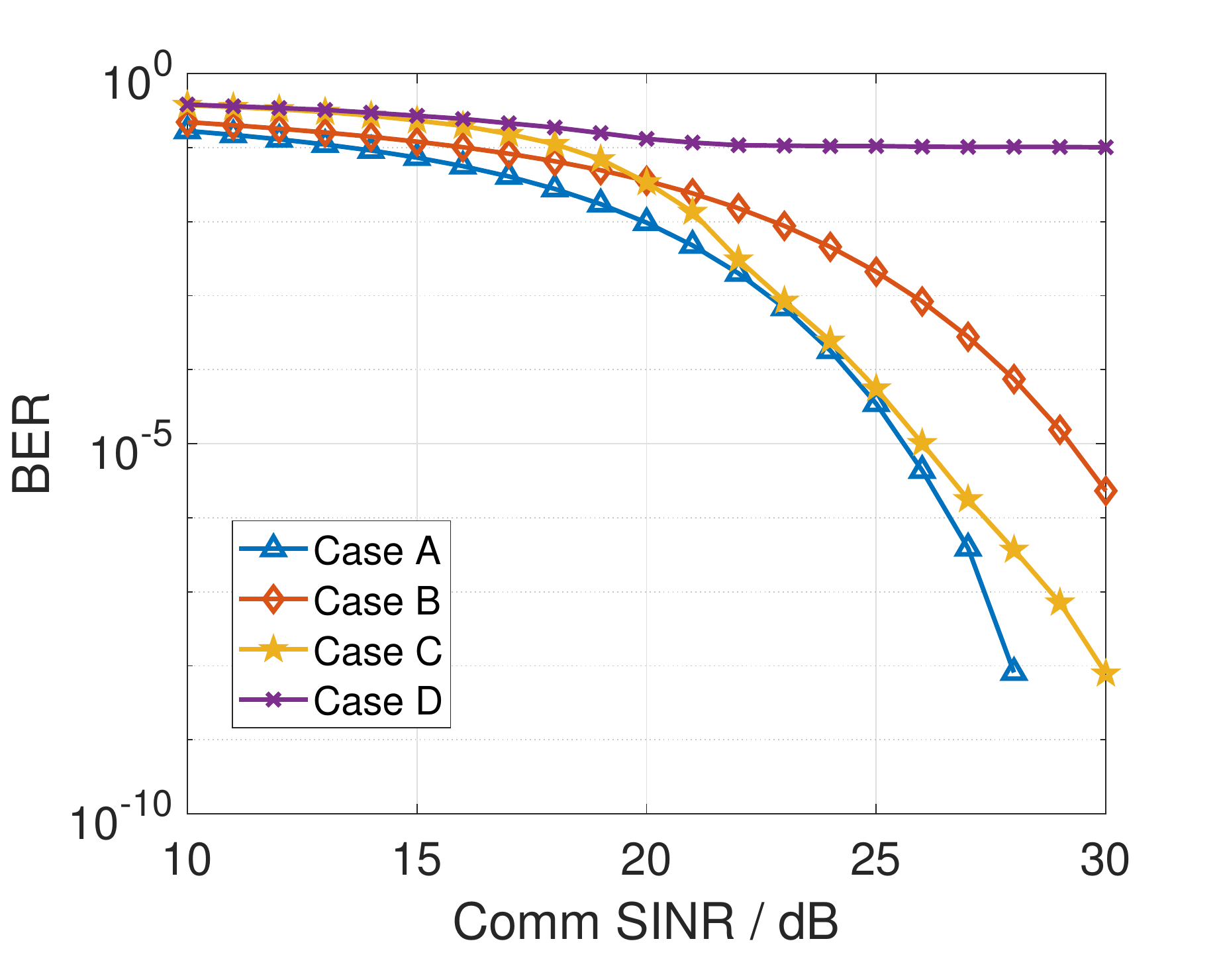}%
	\DeclareGraphicsExtensions.
	\caption{BERs of DL JCS communication}
	\label{fig: BER}
\end{figure}

{\color{blue} 
\subsection{Communication Performance}
We first present the BERs of demodulating communication signals using the CSI obtained by the JCS CSI enhancement method, compared with using the original CSI. For the simplicity of description, we predefine 4 cases for comparison: \textit{Cases} A and B are for demodulating communication signals using the perfect CSI and original estimated CSI, respectively. \textit{Cases} C and D are for demodulating communication signals with the CSI enhanced by the MUSIC-based JCS sensing results and the CSI processed with FFT-based JCS sensing results, respectively. 

Fig.~\ref{fig: BER} shows the BER results when 64-QAM is used for communication. Note that when the detected target is the communication user, the relation between C-SINR and S-SINR is $\frac{{{\gamma _{C,n,m}}}}{{{\gamma _{S,n,m}}}} = \frac{{\left\| {{h_{C,n,m}}} \right\|_2^2}}{{\left\| {{h_{S,n,m,l}}} \right\|_2^2}}$, according to \eqref{equ:gamma_c} and \eqref{equ:gamma_s} under the assumption $\gamma _C^{IN} = \gamma _S^{IN}$ = 3 dB. Due to the CSI estimation error caused by noise and interference, the BER for \textit{case} B is significantly larger than that for \textit{case} A. As C-SINR increases, the BER for \textit{case} C decreases rapidly and becomes lower than that for \textit{case} B after C-SINR is larger than 20 dB. This is because the JCS CSI enhancement method exploits the accurate sensing results and improves the estimated CSI. By comparing \textit{case} D with \textit{cases} B and C, we can see that the BER for \textit{case} D is much larger, which indicates that the FFT-based JCS sensing results are not helpful for improving the CSI and for communication. Referring to the sensing MSEs of the MUSIC-based and the FFT-based JCS in Fig.~\ref{figs:Range_Ms64_Nc128}, we can see that the more accurate the sensing results are, the better CSI enhancement performance is, as the accuracy of range estimation directly determines the accuracy of $A$ in \textbf{Algorithm~\ref{DL_Kalman_CSI}}. This is also the reason that the BER for \textit{case} C decreases rapidly when the sensing MSE becomes sufficiently low.
}

\section{Conclusion}\label{sec:conclusion}
{\color{blue}
In this paper, we proposed a novel JCS system that can achieve accurate AoA, range, and velocity estimation based on improved MUSIC algorithms, together with improved communication performance. Compared with the conventional FFT-based sensing method, our proposed MUSIC-based sensing method can achieve much higher accuracy in range and radial velocity estimation. The proposed JCS CSI enhancement method exploits the JCS sensing results in the design of a Kalman filter for refining the CSI estimate. It is shown to significantly improve the communication performance at high SNRs, approaching the performance with perfect CSI. Moreover, we derived the theoretical lower bound MSEs for the proposed range and velocity estimators using perturbation analysis. Simulation results demonstrate that the theoretical results match the simulation results well, particularly at higher SNRs.
}

\begin{appendices}
	{\color{blue}
	\section{Derivation of $\alpha_t$} \label{Appendix_alpha_t}
	First, we denote the eigenvalue vector as ${{\bf{v}}_x} = \text{diag}({{\bf{\Sigma }}_x})$, where $\text{diag}(\bf X)$ denotes a vector taking the diagonal values of $\bf X$. The mean value of ${{\bf{v}}_x}$ is denoted by ${m_x}$, and ${{\bf{v}}_x} \in \mathbb{R}^{N \times 1}$. We assume there are $L$ incident signals. According to the property of the MUSIC algorithm, the $i$th entry of ${{\bf{v}}_x}$ can be expressed as \cite{MUSIC1986}
	\begin{equation}\label{equ:vx}
		{[ {{{\bf{v}}_x}} ]_i} = \left\{ \begin{array}{l}
			{P_i} + \sigma _N^2, \ i \le L\\
			\sigma _N^2, \ i > L
		\end{array}, \right.
	\end{equation}
	where $P_i$ is the power of the $i$th incident signal, $\sigma _N^2$ is the noise power. We define the differential vector of ${{\bf{v}}_x}$ as ${{\bf{v}}_\Delta }$, where ${[ {{{\bf{v}}_\Delta }} ]_i}{\rm{ = }}{[ {{{\bf{v}}_x}} ]_i} - {[ {{{\bf{v}}_x}} ]_{i + 1}}$, and ${{\bf{v}}_\Delta } \in \mathbb{R}^{( {N - 1} ){\kern 1pt}  \times 1}$, Obviously, ${[ {{{\bf{v}}_\Delta }} ]_i} \approx 0$ when $i > L$, while ${[ {{{\bf{v}}_\Delta }} ]_i} \gg  0$ when $i \le L$. Since mmWave suffers from large propagation loss, $L$ is typically much smaller than $N$. Then, we represent the mean value of the latter half of ${{\bf{v}}_\Delta }$ as $\bar v = {{\sum\limits_{k = \left\lfloor {(N - 1)/2} \right\rfloor }^{N - 1} {{{[ {{{\bf{v}}_\Delta }} ]}_k}} } \mathord{/
			{\vphantom {{\sum\limits_{k = \left\lfloor {(N - 1)/2} \right\rfloor }^{N - 1} {{{[ {{{\bf{v}}_\Delta }} ]}_k}} } {( {N - \left\lfloor {(N - 1)/2} \right\rfloor } )}}} 
			\kern-\nulldelimiterspace} {( {N - \left\lfloor {(N - 1)/2} \right\rfloor } )}}$, and
	$\bar v$ is close to 0. Therefore, the number of detected targets is determined as 
	\begin{equation}
		\hat L = \mathop {\arg \max }\limits_i {[ {{{\bf{v}}_\Delta }} ]_i} > ( {1 + \varepsilon } )\bar v,
	\end{equation}
	where $\varepsilon $ is a parameter used to avoid false detection caused by a small error. In the simulation, we set $\varepsilon = 1$. 
	
	Therefore, ${\alpha _t}$ is set as ${\alpha _t} = {{{{[ {{{\bf{v}}_x}} ]}_{\hat L}}} \mathord{\left/
			{\vphantom {{{{\left[ {{{\bf{v}}_x}} \right]}_{\hat L}}} {{m_x}}}} \right.
			\kern-\nulldelimiterspace} {{m_x}}}$.
	It is a key parameter and has an important impact on sensing accuracy. When ${\alpha _t}$ is too large, the selected noise subspace will include part of the signal subspace, and thus the target may be missed; when ${\alpha _t}$ is too small, the noise subspace is not selected completely, and thus large noise may be taken into the signal space.
	
	\section{Derivation of $\sigma _p^2$} \label{Appendix_sigma_p}
	
	We first derive the eigenvalue matrix of ${\bf{\hat H}}_C{( {{\bf{\hat H}}_C} )^H}$ as ${{\bf{\Sigma }}_p}$, and obtain the eigenvalue vector as ${{\bf{v}}_p} = \text{diag}( {{{\bf{\Sigma }}_p}} )$. When the LoS signal dominates the communication channel, i.e., $L = 1$, from \eqref{equ:vx}, we can estimate $\sigma _p^2$ as 
	\begin{equation}\label{equ:sigma_p_2}
		\hat \sigma _p^2 = {{\sum\limits_{i = 2}^{N_c} {{{[ {{{\bf{v}}_x}} ]}_i}} } \mathord{/
				{\vphantom {{\sum\limits_{i = 2}^{N_c} {{{[ {{{\bf{v}}_p}} ]}_i}} } {(N_c - 1)}}} 
				\kern-\nulldelimiterspace} {(N_c - 1)}}.
	\end{equation}

	}

	\section{Proof of Theorem~\ref{Theo:1}} \label{Theo:A}
			${\bf{U}}_{x,r }$ can be divided as ${\bf{U}}_{x,r } = [ {{\bf{S}}_{x,r },{\bf{U}}_{x,r N}} ]$. Because ${\bf{U}}_{x,r }$ is an orthogonal matrix, ${[ {{\bf{S}}_{x,r }} ]^H}{\bf{U}}_{x,r N}{\rm{ = }}{\bf{0}}$ and ${[ {{\bf{U}}_{x,r N}} ]^H}{\bf{U}}_{x,r N}{\rm{ = }}{\bf{I}}$ hold. 
			
			On one hand, since ${\bf{U}}_{x,r N}$ is the noise subspace of ${\bf{R}}_{{{X}},r }$, we have 
			\begin{equation} \label{equ: Rx_tau_Utau}
					{\bf{R}}_{{{X}},r }{\bf{U}}_{x,r N} = {\sigma _W}^2{\bf{U}}_{x,r N},
			\end{equation}
			where ${\sigma _W}^2$ is the Gaussian noise variance.
			
			On the other hand, we have
			\begin{equation} \label{equ:Rx_tau}
					{\bf{R}}_{{{X}},r } = {{E}}( {{\bf{\bar H}}_S{{[ {\bf{\bar H}}_S ]}^H}} ) \\
					= {{\bf{A}}_{\bf{r}}}E\{ {{\bf{S}}_{r,s}{{[ {{\bf{S}}_{r,s}} ]}^H}} \}{[ {{{\bf{A}}_{\bf{r}}}} ]^H} + {\sigma _W}^2{\bf{I}}
			\end{equation}
			Therefore,
			\begin{equation} \label{equ:Rx_tau_U}
					{\bf{R}}_{{{X}},r }{\bf{U}}_{x,r N} = {{\bf{A}}_{\bf{r}}}E\{ {{\bf{S}}_{r,s}{{[ {{\bf{S}}_{r,s}} ]}^H}} \}{[ {{{\bf{A}}_{\bf{r}}}} ]^H}{\bf{U}}_{x,r N}
					+ {\sigma _W}^2{\bf{U}}_{x,r N}.
			\end{equation}
			By comparing \eqref{equ: Rx_tau_Utau} with \eqref{equ:Rx_tau_U}, we obtain
			\begin{equation} \label{equ:Ar_U_x_N}
				{{\bf{A}}_{\bf{r}}}E\{ {{\bf{S}}_{r,s}{{[ {{\bf{S}}_{r,s}} ]}^H}} \}{[ {{{\bf{A}}_{\bf{r}}}} ]^H}{\bf{U}}_{x,r N} = {\bf{0}}.
			\end{equation}
			Thus,
			\begin{equation} \label{equ:U_x_N_Ar_U_x_N}
				{[ {{\bf{U}}_{x,r N}} ]^H}{{\bf{A}}_{\bf{r}}}E\{ {{\bf{S}}_{r,s}{{[ {{\bf{S}}_{r,s}} ]}^H}} \}{[ {{{\bf{A}}_{\bf{r}}}} ]^H}{\bf{U}}_{x,r N} = {\bf{0}}.
			\end{equation}
			Since $E\{ {{\bf{S}}_{r,s}{{[ {{\bf{S}}_{r,s}} ]}^H}} \}$ is full-rank, ${[ {{\bf{U}}_{x,r N}} ]^H}{{\bf{A}}_{\bf{r}}} = {\bf{0}}$. Therefore, the multiplication between ${[ {{\bf{U}}_{x,r N}} ]^H}$ and each column of ${{\bf{A}}_{\bf{r}}}$ is 0, i.e., ${( {{\bf{U}}_{x,r N}} )^H}{{\bf{a}}_r}( {{r_l}} ) = 0$ holds. Thus, the minimum points of ${\| {{\bf{U}}{{_{x,r N}}^H}{{\bf{a}}_r}( r )} \|_2^2}$ are the ranges.
			
			Similarly, by comparing the two expressions of ${\bf{R}}_{{{X}},f}{\bf{U}}_{x,fN}$, we obtain
			\begin{equation} \label{equ:U_x_N_Af_U_x_N}
				{[ {{\bf{U}}_{x,fN}} ]^H}{{\bf{A}}_{\bf{f}}}E\{ {{\bf{S}}_{f,s}{{[ {{\bf{S}}_{f,s}} ]}^H}} \}{[ {{{\bf{A}}_{\bf{f}}}} ]^H}{\bf{U}}_{x,fN} = {\bf{0}}.
			\end{equation}
			
			Because $E\{ {{\bf{S}}_{f,s}{{[ {{\bf{S}}_{f,s}} ]}^H}} \}$ is full-rank, ${[ {{\bf{U}}_{x,fN}} ]^H}{{\bf{A}}_{\bf{f}}} = {\bf{0}}$ holds. Hence, the multiplication between ${[ {{\bf{U}}_{x,fN}} ]^H}$ and each column of ${{\bf{A}}_{\bf{f}}}$ is 0, i.e., ${( {{\bf{U}}_{x,fN}} )^H}{{\bf{a}}_f}( {{f_{s,l,1}}} ) = 0$ holds. Thus, the minimum points of $\| {({\bf{U}}{{_{x,fN}}})^H{{\bf{a}}_f}( f )} \|_2^2$ are the Doppler results.
			
			The proof of \textbf{Theorem} \ref{Theo:1} is completed.

			\section{} \label{Expressions:G}
			\subsubsection{The derivatives for $\Delta {{\bf{G}}_{\bf{p}}}$}
			The expanded expression for ${{\bf{G}}_p}( {{{\bf{p}}_k};{{{\bf{\tilde U}}}_0}} )$ can be given by
			\begin{equation} \label{equ:delta_p_derivation}
					{{\bf{G}}_p}( {{{\bf{p}}_k};{{{\bf{\tilde U}}}_0}} ) \!=\! 2{\mathop{\rm Re}\nolimits} \left\{ 
						{\bf{a}}_{\bf{p}}^{( 1 )}{( {{{\bf{p}}_k}} )^H}( {{{\bf{U}}_0} + \Delta {{\bf{U}}_0}} )
						\times {( {{{\bf{U}}_0} + \Delta {{\bf{U}}_0}} )^H}{\bf{a}}( {{{\bf{p}}_k}} ) \right\}.
			\end{equation}
			Then, according to \eqref{equ:Gp_expression}, $\Delta {{\bf{G}}_{\bf{p}}}$ can be expressed as 
			\begin{equation} \label{equ:delta_Gp}
				\begin{aligned}
					\Delta {{\bf{G}}_{\bf{p}}} = 2{\mathop{\rm Re}\nolimits} \left\{ {{\bf{a}}_{\bf{p}}^{( 1 )}{{( {{{\bf{p}}_k}} )}^H}\left( \begin{array}{l}
							{{\bf{U}}_0}\Delta {{\bf{U}}_0}^H
							+ \Delta {{\bf{U}}_0}{{\bf{U}}_0}^H
							+ \Delta {{\bf{U}}_0}\Delta {{\bf{U}}_0}^H
						\end{array} \right){\bf{a}}( {{{\bf{p}}_k}} )} \right\}.
				\end{aligned}
			\end{equation}
			By discarding the second-order perturbation $\Delta {{\bf{U}}_0}\Delta {{\bf{U}}_0}^H$ and ${{\bf{U}}_0}^H{\bf{a}}( {{{\bf{p}}_k}} ) = 0$, and substituting \eqref{equ:delta_U0} into \eqref{equ:delta_Gp}, we obtain
			\[\Delta {{\bf{G}}_{\bf{p}}} \!\!=\!\! 2{\mathop{\rm Re}\nolimits}\! \{ \!{ - {\bf{a}}_{\bf{p}}^{( 1 )}{{( {{{\bf{p}}_k}} )}^H}{{\bf{U}}_0}{{\bf{U}}_0}^H{{[ {{\bf{N}}_t} ]}^H}{{\bf{V}}_s}{{\bf{\Sigma }}_s}^{\! - 1}{{\bf{U}}_s}^{\! H} \!{\bf{a}}( {{{\bf{p}}_k}} )\!} \}.\]
			\subsubsection{The derivatives for $\Delta G_r$}
			The expanded expression for $\Delta G_r$ can be given by
			 \begin{equation} \label{equ:delta_Gr_medium}
			 			\Delta G_r = G_r( {r;{{{\bf{\tilde U}}}_{r,0}}} ) - G_r( {r;{{\bf{U}}_{r,0}}} )\\
			 			= 2{\mathop{\rm Re}\nolimits}  [ {{\bf{a}}_r^{( 1 )}{{( r )}^H} \! ( {{{\bf{U}}_{r,0}} \!+\! \Delta {{\bf{U}}_{r,0}}} ){{\! ( {{{\bf{U}}_{r,0}} \!+\! \Delta {{\bf{U}}_{r,0}}} )}^H}\!{{\bf{a}}_r}( r )} \!].
			 \end{equation}
			By discarding the second-order perturbation term, ${{\bf{U}}_{r,0}}^H{{\bf{A}}_{\bf{r}}} = {\bf{0}}$, and substituting \eqref{equ:Ur0} into \eqref{equ:delta_Gr_medium}, we obtain
			\[\Delta G_r \!\! = \!\! {\rm{2Re}}[ {{\bf{a}}_r^{( 1 )}{{\! ( {{r_k}} )}^{\!H}}\!{{\bf{U}}_{r,0}}\!{{\bf{U}}_{r,0}}^{\!H}\! {\bf{W}}_{tr}{{\bf{V}}_{r,s}}{{\bf{\Sigma }}_{r,s}}^{\!\!\!\! - 1}{{\bf{U}}_{r,s}}^{\! \! H}{{\bf{a}}_r}\! ( {{r_k}} )} ].\]	
\end{appendices}


%

{\small
	\bibliographystyle{IEEEtran}
	\bibliography{reference}
}
\vspace{-10 mm}
\ifCLASSOPTIONcaptionsoff
  \newpage
\fi

\end{document}